\documentclass[a4paper,fleqn,usenatbib]{mnras}
\usepackage{newtxtext,newtxmath}
\usepackage[T1]{fontenc}
\usepackage{ae,aecompl}
\usepackage{graphicx}
\usepackage{amsmath}
\usepackage{amssymb}

\title[Fallback accretion in SGRBEEs]{Fallback Accretion onto a Newborn Magnetar: short GRBs with Extended Emission}

\author[S. L. Gibson et al.]{
S. L. Gibson,$^{1}$\thanks{E-mail: slg44@leicester.ac.uk}
G. A. Wynn,$^{1}$
B. P. Gompertz$^{2}$
and P. T. O'Brien$^{1}$
\\
$^{1}$Department of Physics and Astronomy, University of Leicester, University Rd, Leicester LE1 7RH\\
$^{2}$Space Telescope Science Institute, 3700 San Martin Drive, Baltimore, MD 21218
}

\date{Accepted XXX. Received YYY; in original form ZZZ}

\pubyear{2017}

\begin{document}
\label{firstpage}
\pagerange{\pageref{firstpage}--\pageref{lastpage}}
\maketitle

\begin{abstract}
There are a subset of short gamma-ray bursts (SGRBs) which exhibit a rebrightening in their high-energy light curves known as extended emission. These bursts have the potential to discern between various models proposed to describe SGRBs as any model needs to account for extended emission. In this paper, we combine fallback accretion into the magnetar propeller model and investigate the morphological changes fallback accretion has on model light curves and fit to the afterglows of $15$ SGRBs exhibiting extended emission from the \emph{Swift} archive. We have parameterised the fallback in terms of existing parameters within the propeller model and solved for the disc mass and angular frequency of the magnetar over time. We then apply a Markov chain Monte Carlo routine to produce fits to the data. We present fits to our extended emission SGRB sample that are morphologically and energetically consistent with the data provided by \emph{Swift} BAT and XRT telescopes. The parameters derived from these fits are consistent with predictions for magnetar properties and fallback accretion models. Fallback accretion provides a noticeable improvement to the fits of the light curves of SGRBs with extended emission when compared to previous work and could play an important role in explaining features such as variability, flares and long dipole plateaux.
\end{abstract}




\section{Introduction}
\label{sec:intro}

Gamma-ray bursts (GRBs) are the brightest, most intense explosions in the Universe. They are very brief flashes of gamma-rays, lasting from a fraction of a second to several seconds, that occur at a rate of a few per day at random locations throughout the Universe \citep{meszaros06}. GRBs are categorised based on a bimodal distribution in their temporal and spectral properties (e.g.~\citealt{kouveliotou93}): long-soft GRBs and short-hard GRBs (SGRBs). The prompt emission of SGRBs typically lasts $<2$ seconds and their spectra are hard, whereas long-soft GRBs last $>2$ seconds and have softer spectra. However, this $2$ second divide is not strict, e.g.~\citet{bromberg13}, and there is significant overlap between the two distributions including interesting phenomena such as the SGRBs with extended emission (SGRBEEs) discussed in this paper.

SGRBEEs are a subset of SGRBs which show rebrightening in high-energy light curves after the prompt emission spike (approximately $10$ s after trigger), which is referred to as the extended emission (EE; \citealt{norris06}). The peak flux of EE is usually lower than the initial spike but it can last for a few hundred seconds, therefore the total fluence is often higher \citep{perley09}. They are believed to be a subset of SGRBs due to their hard spectra, association with galaxies with low star-forming rates and the lack of any detectable supernovae coincident with the burst. These bursts are an interesting subset to study since any model hoping to describe SGRBs generally needs to account for those which exhibit EE and provide an argument as to why some bursts don't, or determine whether EE is just an observational artefact. Also, a model would need to explain EE energetically and account for the similar total energy in the EE and the prompt emission.

Different mechanisms have been suggested to power EE, including magnetar spin-down \citep{metzger08,bucciantini12}, a two-jet solution \citep{barkov11}, fallback accretion \citep{rosswog07}, $r$-process heating of the accretion disc \citep{metzger10}, and magnetic reconnection and turbulence \citep{zhang11}. Previously, \citet{gompertz14} have implemented a propeller model with a magnetar central engine as an explanation for extended emission bursts. The magnetar is believed to be formed during the merger of two compact objects, i.e. a neutron star binary \citep{rosswog03,belczynski06}, a white dwarf binary \citep{chapman07}, or a neutron star-white dwarf binary. Compact object binary mergers are also the most popular candidates for SGRB progenitors. Magnetars have proven to be a favourable central engine choice since the energy released from their magnetic field via dipole spin-down is comparable to the energy contained within EE. The magnetic propeller model aims to extract the energy required for EE from mass ejected from the system via the propeller mechanism. The version presented in \citet{gompertz14} consists of a static disc which is fully formed at $t=0$ and is drained via either accretion or propellering. The results presented in \citet{gompertz14} run out of energy before fitting the fading afterglow, since the energy reservoir is not replenished, and does not fit to the prompt emission.


Models such as \citet{rosswog07}, \citet{kumar08}, and \citet{cannizzo11} predict the fallback of mass into a disc and so the version of the propeller model presented here has been extended to include fallback accretion. This replenishes the disc and thereby increases the overall available energy budget within the model. This means that the mass of the disc can vary over time as opposed to the static disc presented in \citet{gompertz14} and affects the spin-up of the magnetar thereby changing the morphology of the light curves produced. This extension to the model will allow us to fit the prompt emission and retain enough energy to fit the fading afterglow where previous models could not. The fallback rate is modelled with a $t^{-5/3}$ profile \citep{rosswog07} and the fallback timescale, along with the available fallback mass, have been parameterised in terms of pre-existing parameters within the model. We aim to investigate the morphological changes that fallback introduces into the light curves and to explain the prompt emission (and hence all of the high-energy light curve) with a single model. As well as the addition of fallback mass and disc physics into the model, we have also introduced a new model for the propeller, fitted with variable efficiency parameters, and fitted to prompt emission data which were not included in \citet{gompertz14}.

In Section~\ref{sec:model}, the mathematical theory of the propeller model is presented including: a discussion of significant changes applied for this paper, an exploration of the parameter space and a comparison with previous work by \citet{gompertz14}. Section~\ref{sec:fitsgrbs} introduces the sample of SRGBEEs to be studied and Section~\ref{sec:fitting} describes the method used to fit the model to the data. Discussed results and concluding remarks are presented in  Sections~\ref{sec:results} and~\ref{sec:concs} respectively.

\section{Model Development}
\label{sec:model}

Within the propeller model, the propeller regime is defined according to the relationship between the Alfv\'en radius (the radius at which the dynamics of the gas within the disc is strongly influenced by the magnetic field, $r_{\rm m}$) and the co-rotation radius (the radius at which material in the disc orbits at the same rate as the magnetar surface, $r_{\rm c}$). These radii are defined as follows
\begin{equation}
\label{eg:alfven}
r_{\rm m} = \mu^{4/7} (GM)^{-1/7} {\left( \frac{3 M_{\rm D} (t)}{t_{\nu}} \right)}^{-2/7},
\end{equation}
\begin{equation}
\label{eq:corotation}
r_{\rm c} = (GM / \omega^{2})^{1/3},
\end{equation}
where $\mu$ is the magnetic dipole moment of the central engine, $G$ is the gravitational constant, $M$ is the mass of the central engine, $M_{\rm D}(t)$ is the disc mass at any given time, $\omega$ is the angular frequency of the central engine, and $t_{\nu}$ is the viscous timescale which is given by $t_{\rm \nu}=R_{\rm D}/\alpha c_{\rm s}$. Here $R_{\rm D}$ is the disc radius, $\alpha$ is a viscosity prescription and $c_{\rm s}$ is the sound speed in the disc. We have used $\alpha=0.1$ and $c_{\rm s} = 10^7$ cm/s throughout this work, in keeping with \citet{gompertz14}.

When $r_{\rm c} > r_{\rm m}$, the accretion disc is rotating more rapidly than the magnetic field (assuming the magnetic field rotates rigidly with the magnetar surface) and magnetic torques act to slow the infalling material down and allow it to accrete. In this case, the magnetar gains angular momentum and spins up hence the rotation of the field increases. Conversely if $r_{\rm c} < r_{\rm m}$, the magnetic field is rotating faster than the material and the result is that particles are accelerated to super-Keplerian velocities and ejected from the system. The magnetar loses angular momentum to the ejected material and its rotation is slowed. This is the propeller regime. To prevent the ejected material from exceeding the speed of light, $r_{\rm m}$ is capped at a fraction of the light cylinder radius $r_{\rm lc}$, which is the radius at which the magnetic field lines rotate at the speed of light in order to maintain rigid rotation with the stellar surface. It is difficult to determine where effective coupling between the magnetic field and the plasma breaks down. We have therefore used a conservative estimate of $r_{\rm m}=0.9r_{\rm lc}$ in common with \citet{gompertz14} which allows comparison with their results.

The theory behind the magnetic propeller model is largely based on that presented in \citet{piro11} and \citet{gompertz14}. Therefore, a full description of the model equations will not be presented here and the focus will remain on the amendments required to model fallback accretion. We have assumed the accretion disc has a surrounding mass budget available to fallback smoothly onto the outer radius of the disc on a ballistic timescale of $t^{-5/3}$, in line with models such as \citet{rosswog07}, and mass flows from the inner disc towards the magnetar with an exponential profile.

The radii $r_{\rm m}$ and $r_{\rm c}$ are dependent on the mass of the accretion disc and the rotation frequency of the magnetar. We have modelled the change in disc mass and frequency with the following equations:
\begin{equation}
\label{eq:mdotdisc}
\dot{M}_{\rm D} = \dot{M}_{\rm fb} - \dot{M}_{\rm prop} - \dot{M}_{\rm acc},
\end{equation}
\begin{equation}
\label{eq:omegadot}
\dot{\omega} = \frac{N_{\rm acc} + N_{\rm dip}}{I}.
\end{equation}

\begin{figure}
\includegraphics[width=\columnwidth]{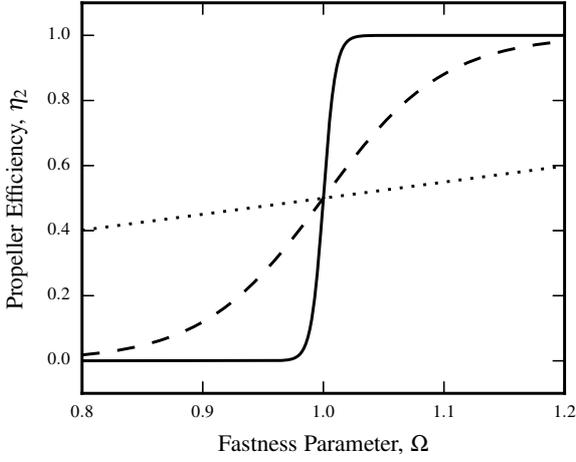}
\caption{A demonstration of how quickly the propeller switches on as a function of $n$ described by Equation (\ref{eq:eta2}). The dotted line corresponds to $n = 1$, the dashed line corresponds to $n = 10$, and the solid line corresponds to $n = 100$.}
\label{fig:tanh}
\end{figure}

Equation~(\ref{eq:mdotdisc}) accounts for mass added to the disc through fallback accretion ($\dot{M}_{\rm fb}$), and mass lost from the disc via the propeller mechanism or accretion onto the magnetar ($\dot{M}_{\rm prop}$ and $\dot{M}_{\rm acc}$ respectively). In Equation~(\ref{eq:omegadot}), $I = 0.35MR^2$ is the magnetar's moment of inertia and $N_{\rm acc}$ and $N_{\rm dip}$ are the accretion and dipole torques acting on the magnetar, respectively. In this work, we adopt the classical dipole torque experienced by any rotating, magnetised body \citep{compactobjects}. $N_{\rm acc}$ has 2 forms dependent on the relationship between $r_{\rm m}$ and the magnetar radius, $R$. If $r_{\rm m} > R$,
\begin{equation}
\label{eq:nacc1}
N_{\rm acc} = {\left(GMr_{\rm m}\right)}^{1/2} \left(\dot{M}_{\rm acc} - \dot{M}_{\rm prop}\right),
\end{equation}
or if $r_{\rm m} < R$,
\begin{equation}
\label{eq:nacc2}
N_{\rm acc} = {\left(GMR\right)}^{1/2} \left(\dot{M}_{\rm acc} - \dot{M}_{\rm prop}\right).
\end{equation}
In the above equations, $\dot{M}_{\rm fb}$, $\dot{M}_{\rm prop}$ and $\dot{M}_{\rm acc}$ are defined as follows.
\begin{equation}
\label{eq:mdotfb}
\dot{M}_{\rm fb} = \frac{M_{\rm fb}}{t_{\rm fb}} \left({\frac{t + t_{\rm fb}}{t_{\rm fb}}}\right)^{-5/3},
\end{equation}
where $M_{\rm fb}$ is the available fallback mass and $t_{\rm fb}$ is the fallback timescale.
\begin{equation}
\label{eq:mdotprop}
\dot{M}_{\rm prop} = \eta_2 \left( \frac{M_{\rm D}(t)}{t_{\rm \nu}} \right),
\end{equation}
\begin{equation}
\label{eq:mdotacc}
\dot{M}_{\rm acc} = (1 - \eta_{\rm 2}) \left( \frac{M_{\rm D}(t)}{t_{\rm \nu}} \right),
\end{equation}
where $\eta_{\rm 2}$ is the efficiency of the propeller mechanism which we define as:
\begin{equation}
\label{eq:eta2}
\eta_{\rm 2} = \frac{1}{2} \Big(1 + \tanh \big[ n ( \Omega - 1 ) \big] \Big).
\end{equation}
This definition of $\eta_{2}$ allows accretion to be turned off at a variable rate as the propeller switches on and the combined efficiency of these mechanisms can never exceed 100\%. In Equation~(\ref{eq:eta2}), $\Omega$ is the `fastness parameter', $\Omega = \omega / (GM_{\rm *}/r_{\rm m}^3)^{1/2} = (r_{\rm m}/r_{\rm c})^{3/2}$, which switches the propeller on as $\Omega \rightarrow 1$, and $n$ controls how ``sharp'' the propeller switch-on is, as demonstrated in Fig.~\ref{fig:tanh}.

\begin{figure}
\includegraphics[width=\columnwidth]{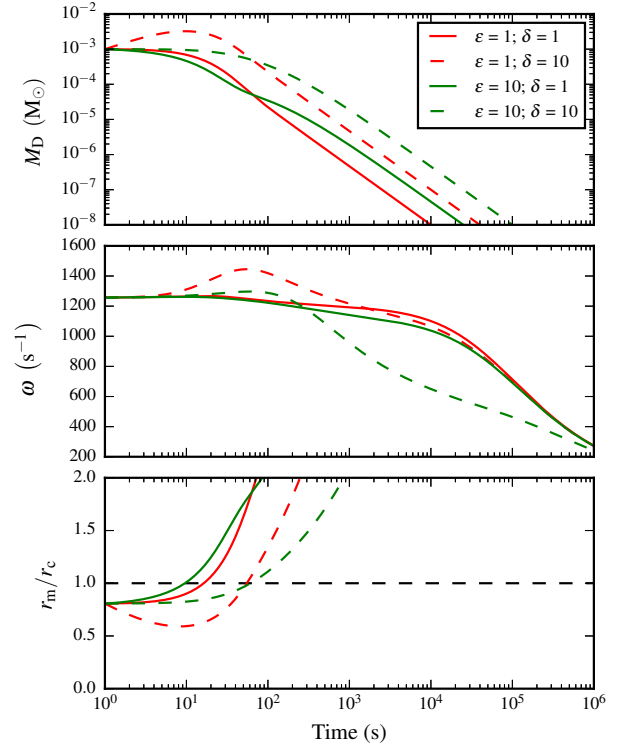}
\caption{A demonstration of how different combinations of the fallback parameters $\epsilon$ and $\delta$ affect the disc mass (top panel) and rotation frequency (centre panel) of a magnetar and disc system with fixed magnetic field, initial spin period, initial disc mass and radius. The bottom panel shows the evolution of the propeller condition $r_{\rm m}/r_{\rm c}$ over time for each combination. The system is in the propeller regime when $r_{\rm m}/r_{\rm c}>1$ (i.e.~above the black, dashed line).}
\label{fig:Momegavst}
\end{figure}

We parameterise the available fallback mass as a fraction ($\delta$) of the initial disc mass, $M_{\rm fb} = \delta M_{\rm D,i}$, and the fallback timescale is similarly parameterised as a fraction ($\epsilon$) of the viscous timescale, $t_{\rm fb} = \epsilon t_{\rm \nu}$. Equations~(\ref{eq:mdotdisc}) and~(\ref{eq:omegadot}) are coupled, first order, ordinary differential equations (ODEs) and, using an ODE integrator, the values of $M_{\rm D}$ and $\omega$ can be calculated for a given range of time points. Fig.~\ref{fig:Momegavst} demonstrates how these fallback parameters affect the disc mass and rotational frequency of a magnetar and disc system and how the propeller condition $r_{\rm m}/r_{\rm c}$ evolves with time.

For short timescales and small fallback masses ($\epsilon = 1$; $\delta = 1$; solid, red curve), the magnetar spins up more slowly despite rapid fallback because the disc is only being fed small amounts of mass. Hence, the propeller mechanism turns on earlier since the propeller condition is at a lower frequency. For short timescales and large fallback masses ($\epsilon = 1$; $\delta = 10$; dashed, red curve), mass is quickly added to the disc and the magnetar spins up rapidly. The propeller mechanism is turned on later because the conditional frequency is higher. For long timescales and small fallback masses ($\epsilon = 10$; $\delta = 1$; solid, green curve), the disc is fed a small amount of mass very slowly and so the magnetar spins up gradually. Again, the propeller condition is at a lower frequency and, therefore, the mechanism turns on earlier. For long timescales and large fallback masses ($\epsilon = 10$; $\delta = 10$; dashed, green curve), the disc mass stays constant over a longer period providing a gentle spin-up of the magnetar. Again, the propeller condition is a higher frequency and the mechanism turns on later. Generally speaking, an initially denser disc makes the propeller mechanism harder to initiate, but the magnetar is spun up more rapidly and, therefore, satisfies the propeller condition at an earlier time.

Once Equations~(\ref{eq:mdotdisc}) and~(\ref{eq:omegadot}) have been integrated, they are then used to estimate the luminosities from the dipole and propelled components, such that
\begin{equation}
\label{eq:Lprop}
L_{\rm prop} = \eta_{\rm prop} \Bigg[ -N_{\rm acc} \omega - \left( \eta_{\rm 2} \frac{GM M_{\rm D}}{r_{\rm m} t_{\rm \nu}}\right) \Bigg]
\end{equation}
and
\begin{equation}
\label{eq:Ldip}
L_{\rm dip} = \eta_{\rm dip} \frac{\mu^2\omega^4}{6c^3},
\end{equation}
where $\eta_{\rm prop}$ and  $\eta_{\rm dip}$ are the propeller and dipole energy-luminosity conversion efficiencies respectively. The total luminosity is given by the sum of the dipole and propeller luminosities and divided by a beaming fraction to account for the relativistic beaming of the jet: $L_{\rm tot} = \left(1/f_{\rm B}\right) \left(L_{\rm dip} + L_{\rm prop}\right)$. $1/f_{\rm B}$ is the fraction of the stellar sphere which is emitting and is related to the half-opening angle of the jet, $\theta_{\rm j}$, as: $1/f_{\rm B} = 1 - \cos (\theta_{\rm j})$ \citep{rhoads99,sari99}.

\subsection{Comparing dipole torque equations}
\label{subsec:torques}

For the dipole torque, we have used the classical solution as given by \citet{compactobjects} and \citet{piro11}.
\begin{equation}
\label{eq:ndip}
N_{\rm dip} = -\frac{\mu^2 \omega^3}{6c^3}.
\end{equation}
The negative sign indicates that $N_{\rm dip}$ spins the magnetar down and produces dipole emission. However, work done by \citet{gompertz14} instead uses the following form for the dipole torque
\begin{equation}
\label{eq:bucc}
N_{\rm dip} = - \frac{2}{3} \frac{\mu^2 \omega^3}{c^3} {\left(\frac{r_{\rm lc}}{r_{\rm m}} \right)}^3,
\end{equation}
which is Equation $(2)$ in \citet{bucciantini06}.

\begin{figure}
\includegraphics[width=\columnwidth]{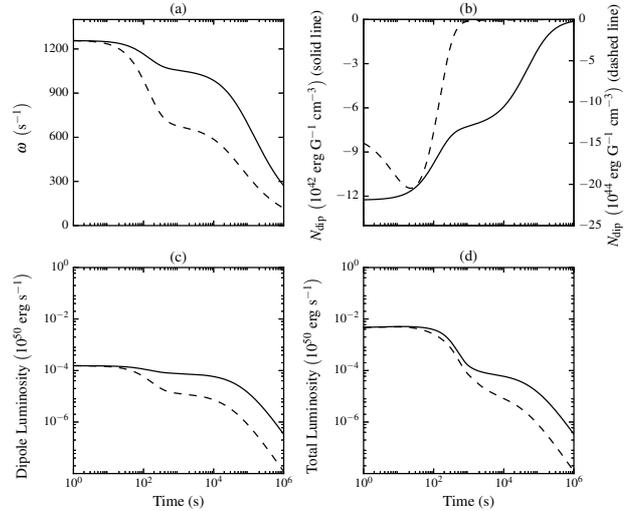}
\caption{A comparison of how Equation~(\ref{eq:ndip}) (solid line; \citealt{piro11}) and Equation~(\ref{eq:bucc}) (dashed line; \citealt{bucciantini06}) affect (a) the stellar spin; (b) the dipole torque; (c) the dipole luminosity; and (d) the total luminosity of a synthetic GRB light curve.}
\label{fig:dipole}
\end{figure}

\citet{bucciantini06} use a relativistic magneto-hydrodynamic (MHD) treatment to solve for the plasma winds emanating from a rotating NS and accretion disc system. They assume that the flow emerges from open flux tubes (providing the extent and shape of the open field line region in the magnetic field is known) and that a truncation of the disc produces more open flux tubes and, therefore, a greater mass loss. Equation~(\ref{eq:bucc}) is then derived from these assumptions. However, it is not certain that these assumptions apply within the model presented in this work and a full MHD treatment of the magnetic propeller is not presented. Therefore, Equation~(\ref{eq:ndip}) is used rather than introduce uncertain assumptions into the model. A comparison between Equations~(\ref{eq:ndip}) and~(\ref{eq:bucc}) is shown in Fig.~\ref{fig:dipole} using a synthetic GRB light curve with arbitrary parameters.

\subsection{Exploring parameter space}
\label{subsec:parameters}

\begin{table}
\centering
\caption{Values used to test the effect of parameter variation on the shape of a GRB light curve. $B$ - magnetic field; $P_{\rm i}$ - initial spin period; $M_{\rm D,i}$ - initial disc mass; $R_{\rm D}$ - disc radius; $\epsilon$ - timescale ratio; $\delta$ - fraction of initial disc mass available in the global mass budget; $n$ - sharpness of propeller switch-on.}
\label{tab:params}
\begin{tabular}{lllllll}
\hline
$B$ & ($10^{15}$ G) & $1$ & $5$ & $10$ & $50$ & - \\
$P_{\rm i}$ & (ms) & $1$ & $5$ & $10$ & - & - \\
$M_{\rm D,i}$ & $\left(M_{\rm \odot}\right)$ & $10^{-5}$ & $10^{-4}$ & $10^{-3}$ &
$10^{-2}$ & $10^{-1}$ \\
$R_{\rm D}$ & (km) & $100$ & $500$ & $1000$ & - & - \\
$\epsilon$ & & $1$ & $10$ & - & - & - \\
$\delta$ & & $1$ & $10$ & - & - & -\\
$n$ & & $1$ & $10$ & $50$ & - & - \\
\hline
\end{tabular}
\end{table}

\begin{figure*}
\includegraphics[width=\textwidth]{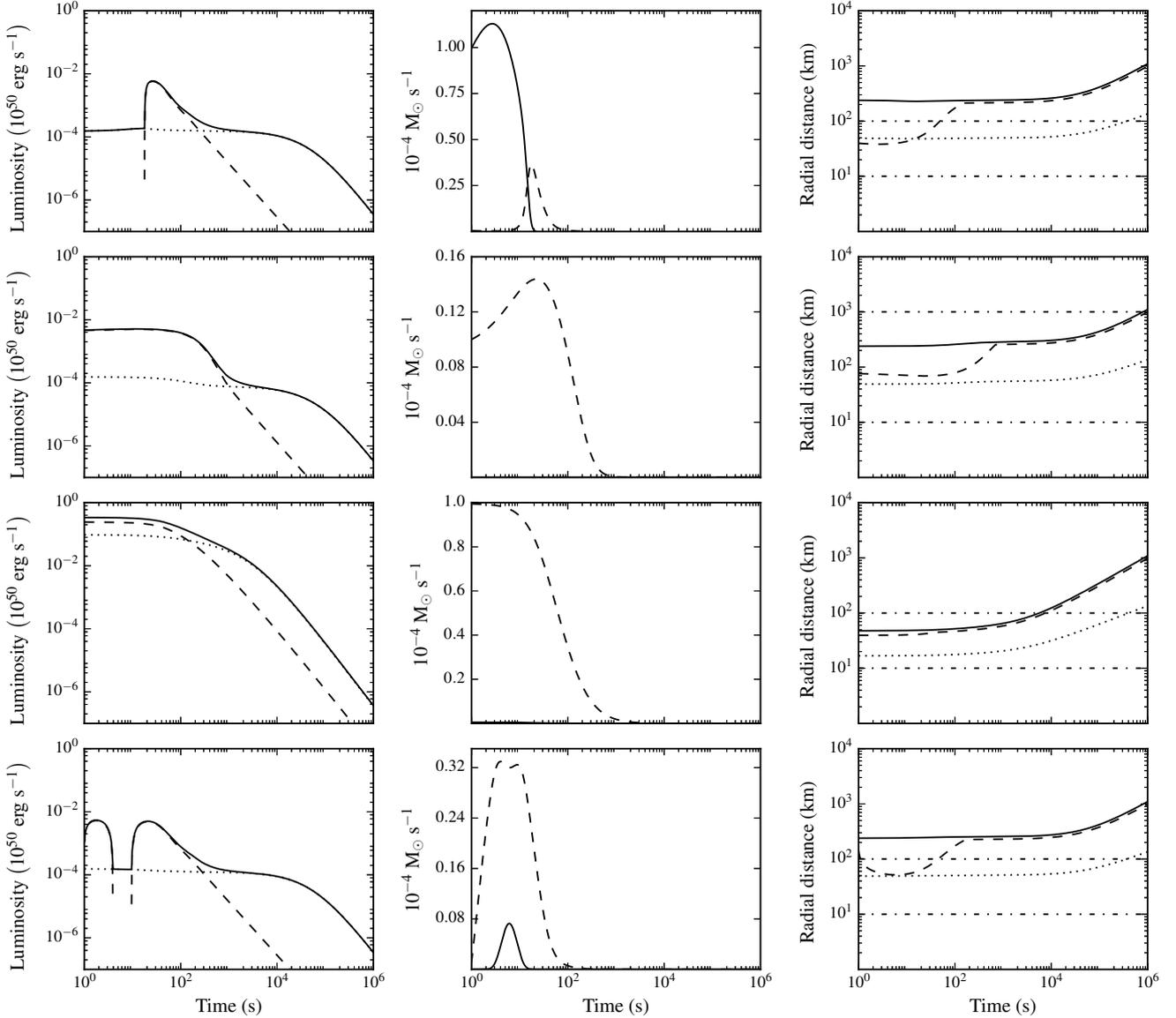}
\caption{Top to bottom: Type I - \emph{Humped}; Type II - \emph{Classic}; Type III - \emph{Sloped}; Type IV - \emph{Stuttering}. Each row shows plots for one example of each class. They are not fully representative of the range of energetics or morphology for their respective classes since they are intended to highlight the light curve shapes only. Left panels: synthetic light curves representing the four phenomenological classes. Dotted line - dipole luminosity; dashed line - propeller luminosity; solid line - total luminosity. Centre panels: mass flow rates in the system. Solid line - mass flow rate on to the central magnetar; dashed line - propellered mass flow out of the system. Right panels: positions of key radii relative to the centre of the magnetar. Dashed line - Alfv\'en radius; dotted line - co-rotation radius; solid line - light cylinder radius. Lower horizontal dot-dashed line is the magnetar radius, upper horizontal dot-dashed line is the outer disc radius, $R_{\rm D}$.}
\label{fig:types}
\end{figure*}

To determine how the modifications to the propeller model have affected the phenomenological classes outlined in \citet{gompertz14} (\emph{humped}, \emph{classic}, \emph{sloped} and \emph{stuttering}), the parameter variation experiment they originally performed was repeated with values from Table~\ref{tab:params}. The magnetar mass and radius were fixed to be $1.4~M_{\rm \odot}$ and $10$ km respectively, the propeller and dipole efficiencies were set to $100\%$ and the the beaming fraction to $1$ since they only act to normalise the luminosity here. The produced light curves represented all combinations of $B$, $P_{\rm i}$, $M_{\rm D,i}$, $R_{\rm D}$, $\epsilon$, $\delta$, and $n$. The four phenomenological types originally outlined in \citet{gompertz14} were recovered and examples of each are shown in Fig.~\ref{fig:types}. All values for $n$ appeared commonly in each type suggesting that the model is insensitive to $n$.

\subsection{Comparing types to previous work}
\label{subsec:comparetypes}

\begin{table}
\centering
\caption{Main parameters used to compare light curves from the previous model (\citealt{gompertz14}) with the modified model without fallback accretion.}
\label{tab:models}
\begin{tabular}{llllll}
\hline
& & Humped & Classic & Sloped & Stuttering \\
\hline
$B$ & ($10^{15}$ G) & $1$ & $1$ & $10$ & $5$ \\
$P_{\rm i}$ & (ms) & $5$ & $5$ & $5$ & $5$ \\
$M_{\rm D,i}$ & $\left(M_{\rm \odot}\right)$ & $10^{-3}$ & $10^{-4}$ & $10^{-4}$ &
$10^{-2}$ \\
$R_{\rm D}$ & (km) & $100$ & $1000$ & $1000$ & $500$ \\
\hline
\end{tabular}
\end{table}

\begin{figure}
\includegraphics[width=\columnwidth]{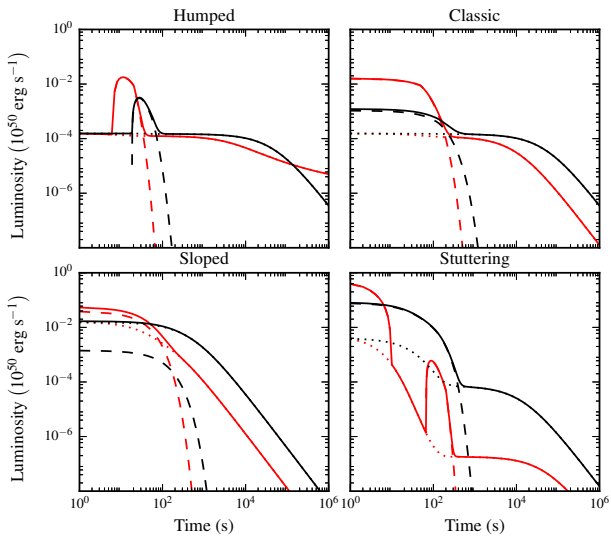}
\caption{Comparison of light curves generated by the previous model (\citealt{gompertz14}; red curves) and the modified model without fallback accretion (black curves). The fallback was turned off by setting $\epsilon=1$ and $\delta=10^{-6}$; $n=1$ as the closest approximation to the switch on in the previous model. Solid lines - total luminosity; dashed lines - propeller luminosity; dotted lines - dipole luminosity.}
\label{fig:comparetypes}
\end{figure}

In order to determine how well the modified model recovered the four types, the parameters given in Table~\ref{tab:models} were used to generate light curves using the previous model described in \cite{gompertz14}. The fallback accretion in the modified model was turned off by setting $\epsilon=1$ and $\delta=10^{-6}$, i.e. the amount of fallback mass is so negligible that the magnetar behaves as if only the accretion disc is present and the fallback timescale becomes irrelevant. The value of $n$ used was $1$ as this is the closest approximation to the propeller switch-on modelled in previous work. Fig.~\ref{fig:comparetypes} compares the modified model without fallback to the previous work. The difference in dipole luminosity between the two models is explained by our use of the classical dipole torque as discussed in Section~\ref{subsec:torques}. Equation~(\ref{eq:ndip}) has a longer dipole duration than Equation~(\ref{eq:bucc}) causing some morphological differences. However, the modified model does not recover the propeller luminosity in all cases, the stuttering type being the most different. Since we have already seen in Fig.~\ref{fig:types} that the modified model is capable of reproducing all types successfully, it is suggested that they have moved in parameter space due to the inclusion of $\dot{M}_{\rm prop}$ and it's link to $\dot{M}_{\rm acc}$ through $\eta_{\rm 2}$.

\section{\emph{Swift} SGRBEE Sample}
\label{sec:fitsgrbs}

The data for the GRB sample were collected by \emph{Swift}. The \emph{Swift} satellite \citep{gehrels04}, launched in 2004, is a multi-wavelength observatory dedicated to GRB hunting with rapid slewing capabilities. It carries three instruments: the Burst Alert Telescope (BAT; \citealt{barthelmy05a}), the X-ray Telescope (XRT; \citealt{burrows05}), and the Ultra-Violet/Optical Telescope (UVOT; \citealt{roming05}). The \emph{Swift} mission and the UK \emph{Swift} Science Data Centre (UKSSDC\footnote{\label{footnote:swift}www.swift.ac.uk}, \citealt{evans07,evans09}) provided the data presented in this paper.

The data need to undergo a cosmological $k$-correction and absorption correction, as described in \citet{bloom01}, to produce bolometric (1 - 10000 keV), redshift-corrected light curves before they can be fitted by the model. This method requires the photon index, $\Gamma$, the absorption coefficient, $\sigma$ (given by the ratio of counts-to-flux unabsorbed to counts-to-flux observed, which are all available on the UKSDCC repository) and the redshift, $z$, some of which were found in the literature (see Table~\ref{tab:sgrbs}). For those GRBs with no measured redshift, the sample mean of $0.39$ from \citet{gompertz14} was used. Alternatively, a randomly generated redshift (e.g.~within $1,~2$ or even $3$ standard deviations of the mean value) could be used. The effect of an increasing $z$ is an increase in luminosity and earlier on-set times that, as we will see later in this paper, causes the model to favour larger initial disc masses and fallback mass budgets. Since these may not have a physical basis, we have chosen to use the sample mean, as in previous work by \citet{gompertz14}.

\begin{table}
\centering
\caption{The sample of SGRBEEs and the parameters required for a cosmological $k$-correction. For GRBs with an unknown redshift (marked with an $^*$), the sample mean of $0.39$ from \citet{gompertz14} was used. $^{\dagger}$Upper limit \citep{davanzo09}. $^{\rm a}$\citet{prochaska05}; $^{\rm b}$\citet{soderberg05}; $^{\rm c}$\citet{price06b}; $^{\rm c}$\citet{berger07}; $^{\rm d}$\citet{cenko06}; $^{\rm e}$\citet{graham09}; $^{\rm f}$\citet{davanzo07}; $^{\rm g}$\citet{selsing16}.}
\label{tab:sgrbs}
\begin{tabular}{llll}
\hline
GRB & $\Gamma$ & $\sigma$ & $z$ \\
\hline
050724 & $1.58^{+0.21}_{-0.19}$ & $1.26$ & $0.2578^{\rm a}$ \\ [2pt]
051016B & $1.85^{+0.14}_{-0.13}$ & $1.31$ & $0.9364^{\rm b}$ \\ [2pt]
051227 & $2.1^{+0.4}_{-0.4}$ & $1.31$ & $2.8^{\dagger}$ \\ [2pt]
060614 & $1.78^{+0.08}_{-0.08}$ & $1.06$ & $0.1254^{\rm c}$ \\ [2pt]
061006 & $2.1^{+0.6}_{-0.4}$ & $1.61$ & $0.4377^{\rm c}$ \\ [2pt]
061210 & $2.60^{+1.92}_{-0.71}$ & $3.48$ & $0.4095^{\rm d}$ \\ [2pt]
070714B & $1.79^{+0.24}_{-0.22}$ & $1.15$ & $0.9224^{\rm e}$ \\ [2pt]
071227 & $1.5^{+0.6}_{-0.5}$ & $1.02$ & $0.381^{\rm f}$ \\ [2pt]
080123 & $2.46^{+1.04}_{-0.70}$ & $1.71$ & $0.39^*$ \\ [2pt]
080503 & $2.38^{+0.42}_{-0.16}$ & $1.24$ & $0.39^*$ \\ [2pt]
100212A & $1.99^{+0.40}_{-0.18}$ & $1.37$ & $0.39^*$ \\ [2pt]
100522A & $2.40^{+0.17}_{-0.16}$ & $2.45$ & $0.39^*$ \\ [2pt]
111121A & $1.78^{+0.21}_{-0.20}$ & $1.42$ & $0.39^*$ \\ [2pt]
150424A & $1.98^{+0.24}_{-0.22}$ & $1.23$ & $0.39^*$ \\ [2pt]
160410A & $1.5^{+0.7}_{-0.6}$ & $1.02$ & $1.717^{\rm g}$ \\
\hline
\end{tabular}
\end{table}

The sample studied in \citet{gompertz13} and \citet{gompertz14} has been expanded here by selecting identified SGRBEEs from \citet{kaneko15} (which covers bursts to the end of 2012) that have good data available in the \emph{Swift} archive. Plus GRBs 150424A and 160410A which are identified as EE bursts within GCN Circulars (\citealt{norris15} and \citealt{sakamoto16} respectively). The data used in the fitting incorporates XRT data and BAT data that have been extrapolated into the XRT bandpass (available from the UKSDCC Burst Analyser tool) since the effect of the extended emission is not always evident in the XRT light curve alone.

\section{Fitting Routine}
\label{sec:fitting}

A Markov chain Monte Carlo  simulation (MCMC; \citealt[chap.~4]{mackay03}) was used to fit the model to data as there are a minimum of six parameters and the MCMC will efficiently search a large portion of parameter space and increase the probability of finding the global minimum of the model. However, the MCMC method requires a burn-in phase which is loosely defined as an unknown number of steps at the beginning of the simulation where each ``walker'' attempts to find the lowest area of probability space. The chain is generally considered to be burned in when all walkers have converged onto this area of probability space. The ``emcee'' module was used to handle the MCMC \citep{emcee}. To construct the posterior probability distribution, a Gaussian log-likelihood function of the following form was chosen
\begin{equation}
\label{eq:lnlike}
\ln (p_{\rm likelihood}) = -\frac{1}{2} \sum_{i=1}^N {\left( \frac{y_i - \hat{y}_i}{\sigma_i} \right)}^{2},
\end{equation}
where $y_i$ is a data point, $\sigma_i$ is it's associated uncertainty, and $\hat{y}_i$ is a model point calculated at the same $x$-value as $y_i$. The \emph{Swift} light curves used here are binned to contain a minimum of $20$ photons per time bin (an exception may be applicable in the last bin) making Gaussian statistics suitable. A prior probability that is flat when the parameters are within the limits given in Table~\ref{tab:limits} was also chosen.
\begin{equation}
\label{eq:lnprior}
\ln (p_{\rm prior}) = \left\{
  \begin{array}{lr}
    ~0 & : x_{l} < x < x_{u}\\
    - \infty & : otherwise
  \end{array}
\right.
\end{equation}
Hence, the full posterior probability distribution is given by
\begin{equation}
\label{eq:lnpost}
\ln (p) = \ln (p_{\rm likelihood}) + \ln (p_{\rm prior}).
\end{equation}

\begin{table}
\centering
\caption{Upper and lower limits placed on the fitting parameters in the MCMC. $M_{\rm D,i}$, $R_{\rm D}$, $\epsilon$, and $\delta$ were searched in log-space for efficiency.}
\label{tab:limits}
\begin{tabular}{llll}
\hline
& & Lower & Upper \\
\hline
$B$ & ($10^{15}$ G) & $10^{-3}$ & $10$ \\
$P_{\rm i}$ & (ms) & $0.69$ & $10$ \\
$M_{\rm D,i}$ & ($M_{\rm \odot}$) & $10^{-3}$ & $10^{-1}$ \\
$R_{\rm D}$ & (km) & $50$ & $2000$ \\
$\epsilon$ & & $0.1$ & $1000$ \\
$\delta$ & & $10^{-5}$ & $50$ \\
$\eta_{\rm dip}$ & (\%) & $1$ & $100$ \\
$\eta_{\rm prop}$ & (\%) & $1$ & $100$ \\
$1/f_{\rm B}$ & & $1$ & $600$ \\
\hline
\end{tabular}
\end{table}

For the MCMC, $100$ affine invariant walkers \citep{goodman10} were used and ran for a $50,000$ step burn in phase to allow the walkers to test all of parameter space. After this run, the best $100$ distinct probabilities were chosen to serve as the starting point for the final MCMC run of the same length. This made sure that the parameters recovered were representative of the global minimum, not a local minimum, and reduces the burn-in of the chain to $\lesssim 1000$ steps in most cases. Although, if the time series (parameter or probability value vs. model number for each walker) showed that the chain had not fully converged, the process of selecting the $100$ best probabilities was repeated and the chain run again until convergence was achieved. The optimal parameters were found by taking the median of the posterior probability distributions and their uncertainties are given by the $95\%$ percentiles. We chose the median, rather than the mean or mode, since it is less sensitive to the tails of distributions and is preserved under reversible transformations of the data (e.g.~$\log_{10} \epsilon \rightarrow \epsilon$). Fits for the SGRBEE sample were produced with a range of free parameters $(p)$: $p=6$ ($B$, $P_{\rm i}$, $M_{\rm D,i}$, $R_{\rm D}$, $\epsilon$ and $\delta$); $7$ (original $6$ plus $1/f_{\rm B}$); $8$ (original $6$ plus $\eta_{\rm dip}$ and $\eta_{\rm prop}$); and $9$ (all listed parameters). $\eta_{\rm dip}$, $\eta_{\rm prop}$ and $1/f_{\rm B}$ were fixed to $5\%$, $40\%$  and $1$ respectively when they were not free parameters, in keeping with \citet{gompertz14}. The fits were repeated for fixed values of $n=1,~10,~100$ and the corrected Akaike Information Criterion (AICc; \citealt{akaike}) was used to establish the best fitting models. We chose this statistic since it allows us to compare models of varying free parameter number ($p$).

AICc is given by the following equation
\begin{equation}
\label{eq:aicc}
AICc = -2 \ln(L) + 2k + \frac{2k(k + 1)}{N - k - 1},
\end{equation}
where $k$ is the number of free parameters and $N$ is the number of observations in the data set. This penalises a model for `overfitting' and scales with $k$. We have substituted Equation~(\ref{eq:lnlike}) for the maximum log-likelihood $\ln(L)$, which then cancels down to the $\chi^2$ statistic. The minimum AICc value within a set is then representative of the optimum model fit since if the AICc value of a model that has a large number of free parameters (and hence a large penalty) is less than a model with fewer free parameters (and hence a small penalty), then it can be generally assumed that the extra parameters improve the quality of fit.

\section{Results and Discussion}
\label{sec:results}

\begin{table*}
\centering
\caption{AICc values for models using $6$, $7$, $8$ and $9$ free parameters ($p$) for $n=1,10,100$. Underlined values are the lowest AICc values for each $n$ bracket and values in \textbf{bold} are the minima across all values of $n$. Values marked with an $^{*}$ are modified since $N-k-1=0$ for these models.}
\begin{tabular}{lcrrrrcrrrrcrrrr}
\hline
& \vline & \multicolumn{4}{c}{$n=1$} & \vline & \multicolumn{4}{c}{$n=10$} & \vline & \multicolumn{4}{c}{$n=100$} \\
GRB & \vline & $p=6$ & $p=7$ & $p=8$ & $p=9$ & \vline & $p=6$ & $p=7$ & $p=8$ & $p=9$ & \vline & $p=6$ & $p=7$ & $p=8$ & $p=9$ \\
\hline
050724 & \vline & $2,415$ & $1,790$ & $1,975$ & $\underline{1,682}$ & \vline & $2,496$ & $2,216$ & $2,005$ & \boldmath$1,507$ & \vline & $2,509$ & $2,128$ & $2,007$ & $\underline{1,610}$ \\
051016B & \vline & $802$ & $540$ & $771$ & $\underline{531}$ & \vline & $785$ & $562$ & $742$ & $\underline{518}$ & \vline & $785$ & $549$ & $742$ & \boldmath$435$ \\
051227 & \vline & $538$ & $267$ & $318$ & $\underline{235}$ & \vline & $535$ & $271$ & $317$ & $\underline{235}$ & \vline & $535$ & $381$ & $317$ & \boldmath$233$ \\
060614 & \vline & $47,390$ & $47,256$ & $47,521$ & \boldmath$44,709$ & \vline & $48,428$ & $47,275$ & $46,950$ & $\underline{44,746}$ & \vline & $48,422$ & $47,263$ & $48,013$ & $\underline{45,278}$ \\
061006 & \vline & $269$ & $\underline{244}$ & $252$ & $250$ & \vline & $317$ & \boldmath$242$ & $543$ & $253$ & \vline & $323$ & $\underline{244}$ & $262$ & $257$ \\
061210 & \vline & $\underline{673}$ & $802$ & $1,054^*$ & $103,101$ & \vline & $677$ & \boldmath$149$ & $1,649^*$ & $5,148$ & \vline & $670$ & $\underline{210}$ & $969^*$ & $5,127$ \\
070714B & \vline & $1,303$ & $1,419$ & $\underline{1,260}$ & $1,352$ & \vline & $1,302$ & $1,397$ & $\underline{1,260}$ & $1,301$ & \vline & $1,302$ & $1,844$ & $1,260$ & \boldmath$1,180$ \\
071227 & \vline & $335$ & $161$ & $163$ & \boldmath$158$ & \vline & $226$ & $339$ & $\underline{161}$ & $163$ & \vline & $225$ & $265$ & $\underline{161}$ & $230$ \\
080123 & \vline & $308$ & $308$ & $\underline{298}$ & $305$ & \vline & $337$ & $\underline{291}$ & $308$ & $307$ & \vline & $360$ & \boldmath$283$ & $330$ & $360$ \\
080503 & \vline & $\underline{2,335}$ & $2,375$ & $2,474$ & $2,474$ & \vline & $2,375$ & $2,379$ & \boldmath$2,294$ & $2,583$ & \vline & $2,927$ & $\underline{2,475}$ & $3,539$ & $2,784$ \\
100212A & \vline & $9,055$ & $8,372$ & $8,130$ & \boldmath$7,310$ & \vline & $9,299$ & $8,271$ & $8,988$ & $\underline{8,196}$ & \vline & $9,258$& $\underline{8,498}$ & $8,988$ & $8,588$ \\
100522A & \vline & $29,377$ & $\underline{22,992}$ & $27,472$ & $23,419$ & \vline & $27,666$ & $23,326$ & $26,116$ & \boldmath$22,184$ & \vline & $27,531$ & $22,460$ & $26,744$ & $\underline{22,411}$ \\
111121A & \vline & $1,754$ & $\underline{1,747}$ & $1,761$ & $1,803$ & \vline & $1,753$ & $1,751$ & $\underline{1,748}$ & $1,766$ & \vline & $1,753$ & $1,750$ & $1,761$ & \boldmath$1,742$ \\
150424A & \vline & $2,377$ & $2,997$ & $2,223$ & \boldmath$1,334$ & \vline & $2,315$ & $51,246$ & $\underline{2,170}$ & $58,080$ & \vline & $2,315$ & $1,795$ & $2,171$ & $\underline{1,432}$ \\
160410A & \vline & $974$ & $\underline{473}$ & $546$ & $18,742$ & \vline & $1,115$ & $\underline{366}$ & $546$ & $403$ & \vline & $1,255$ & $412$ & $546$ & \boldmath$359$ \\
\hline
\end{tabular}
\label{tab:aicc}
\end{table*}

\begin{figure*}
\includegraphics[width=\textwidth]{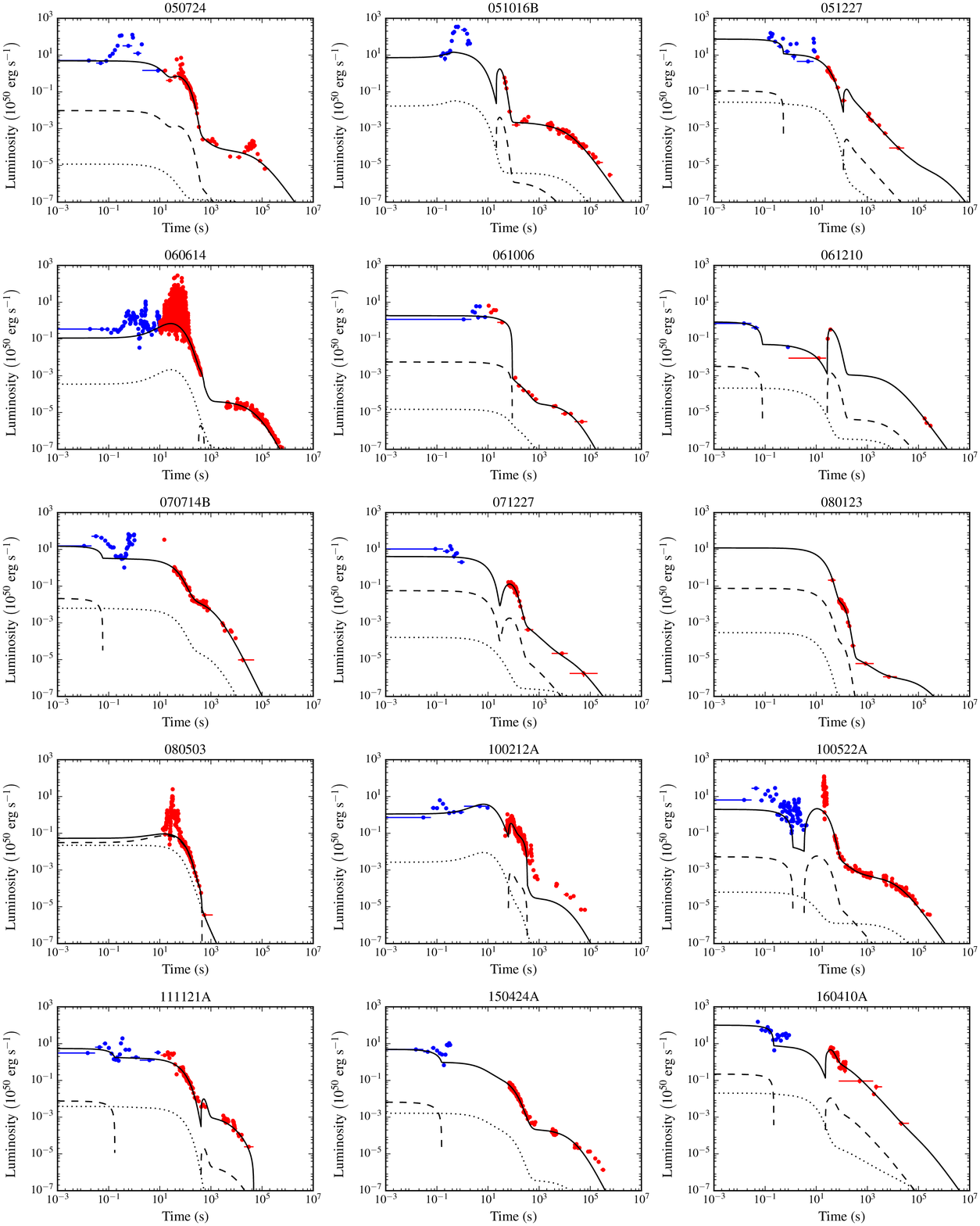}
\caption{Best global fits to the SGRBEE sample (bold values in Table~\ref{tab:aicc}). Dashed line - propeller luminosity; dotted line - dipole luminosity; solid line - total luminosity; red points - combined BAT and XRT data.}
\label{fig:results}
\end{figure*}

Table~\ref{tab:aicc} presents the AICc values for all results of the fitting routine. The large spread of values is representative of the difficulty $\chi^2$ (the root of the AICc) has comparing a smooth model with highly variable data, especially in the early-time BAT data. Table~\ref{tab:aicc} shows that the general picture of the model is stable over all $n$ values since there is a reasonable spread of best fits. This also confirms the observation made in Section~\ref{subsec:parameters} that the model is reasonably insensitive to $n$. Increasing $n$ only makes features such as humps appear sharper, which does not have a great impact on the overall quality of the fit. The best global fits to the SGRBEE sample (bold values in Table~\ref{tab:aicc}) are presented in Fig.~\ref{fig:results}.

The $p=6$ set represents the core physics of the model by constraining the fundamental properties of the magnetar ($B$ and $P_{\rm i}$), the accretion disc ($M_{\rm D,i}$ and $R_{\rm D}$) and the fallback ($M_{\rm fb}$ and $t_{\rm fb}$ through $\delta$ and $\epsilon$ respectively) and is the most energetically restricted case compared to the $p=9$ case which has the largest energy reservoir. Furthermore, $\eta_{\rm dip}$ and $\eta_{\rm prop}$ determine the efficiency at which the dipole and propeller mechanisms respectively need to work at in order to convert the energy to luminosity. Lastly, $f_{\rm B}$ accounts the anisotropy of the radiation ($1/f_{\rm B}$ is the solid angle of emission). The results of the MCMC were analysed for parameter correlations though none were found since our method of selecting the best probabilities after the burn-in phase removes any correlation by placing the parameters in the global minimum.

The $k$-correction performed in Section~\ref{sec:fitsgrbs} assumes isotropic emission, whereas in actuality, GRBs are beamed into a very narrow opening angle due to their relativistic velocity \citep{fruchter99,harrison99,frail01}. Rather than divide the data down to a beam-corrected level, our routine works to multiply the model up to the isotropic luminosity level so that model comparison becomes easier on the same scale. The morphologies of the fits change as each new parameter is introduced since they handle the high luminosities at early times allowing the core parameters to reconfigure. This means that there can be more energy available at late times to fit the fading afterglow.

It is interesting to compare the freedom of the model (i.e.~how many free parameters are used) with the ``sharpness'' of the propeller (i.e.~the $n$ value). Generally speaking, the AICc value of the fit improves as the number of free parameters increases, whereas, increasing $n$ for the same number of free parameters often does not improve the fit.
Also, $p=8$ fits often perform worse than $p=7$ fits implying that the beaming fraction has a greater role within the model than the efficiencies, but the inclusion of all $3$ of these parameters are most preferable. Table~\ref{tab:beaming} shows a comparison of the jet half-opening angles derived from the best fits in this work with hydrodynamical modelling performed by \citet{ryan15} for $4$ GRBs common to both studies. Our model produces systematically narrower jets (most likely caused by the models attempts to fit the early-time luminosity) which are partially consistent with \citet{ryan15} in errors (e.g. GRBs 051016B and 060614), and where they are not (e.g.~GRB 061006), they are broadly consistent to $\sim 2 - 2.5 \sigma$.

\begin{table}
\centering
\caption{Table showing the half-opening angles (in radians) for $4$ GRBs, calculated from $f_{\rm B}=1-\cos\left(\theta_{\rm j}\right)$. $\theta_{\rm j}$ values are from the global best fits of this work (uncertainties are $95\%$ confidence interval); $\theta_{\rm 0}$ values are from \citet{ryan15}.}
\label{tab:beaming}
\begin{tabular}{lcc}
\hline
GRB & $\theta_{\rm j}$ & $\theta_{\rm 0}$ \\
\hline
051016B & $0.07^{+0.11}_{-0.11}$ & $0.35^{+0.11}_{-0.24}$ \\ [2pt]
060614 & $0.079^{+0.359}_{-0.500}$ & $0.293^{+0.122}_{-0.085}$ \\ [2pt]
061006 & $0.078^{+0.088}_{-0.081}$ & $0.407^{+0.068}_{-0.173}$ \\ [2pt]
070714B & $0.06^{+0.18}_{-0.12}$ & $0.33^{+0.11}_{-0.11}$ \\
\hline
\end{tabular}
\end{table}

Comparing our results with that of \citet{gompertz14}, we can see the inclusion of fallback accretion within the propeller model allows for an improvement in fitting the `tail' of the fading afterglow. This is can be seen in GRBs 051227, 060614 and 061006 where \citet{gompertz14} did not produce such good fits to the tail. Hence, fallback accretion is a necessary addition to the propeller model in order to fully explain the energetics and morphologies of SGRBEEs. Additionally, the extended model handles variability and flares within the data much more naturally than \citet{gompertz14} and copes with the early-time luminosity detected by BAT.

The parameters derived from the fits in Fig.~\ref{fig:results} are presented in Table~\ref{tab:results_pars}. We find that the magnetic fields derived from the fits are in the moderate to high end of the parameter space and that the sample generally have slow initial spins. The slow initial spins are most likely due to the additional fallback spinning the magnetar up and, therefore, the constraints on high initial spin rates is relaxed. This has an impact on the value of the magnetic field derived as the fit moves along the correlation between $B$ and $P_{\rm i}$ discussed in \citet{gompertz14}. The sample fits also tend to favour massive discs and narrow jet opening angles. This is most likely due to the model extracting as much of the available energy as possible to fit the high luminosities at early times in the light curve, data which was not included in the fits of \citet{gompertz14}. The values of $\epsilon$, $\delta$, $\eta_{\rm dip}$ and $\eta_{\rm prop}$ are widely distributed throughout the parameter space. The derived parameters are consistent with predictions for a magnetar (\citealt{giacomazzo13}; \citealt{mereghetti15}; \citealt{rea15}) and are also consistent with the results in \citet{gompertz14}.

We will now examine how increasing the number of free parameters affects the fits in $3$ GRBs from the sample. GRB 060614 has been chosen since this is a uniquely interesting burst given its characteristics. GRBs 050724 and 111121A were chosen as examples of the model behaving consistently well, or vice versa, over the different parameter sets.

\begin{table*}
\centering
\caption{Parameters derived from the best global fits to the SGRBEE sample (bold values in Table~\ref{tab:aicc}). Reported errors are $95\%$. Values marked with an [L] have reached a parameter limit; those marked with [F] were fixed during fitting. The $\chi^{2}_{\rm red}$ values are also presented to indicate goodness of fit.}
\label{tab:results_pars}
\begin{tabular}{lccccccccccc}
\hline
GRB & $n$[F] & $B$ & $P_{\rm i}$ & $M_{\rm D,i}$ & $R_{\rm D}$ & $\epsilon$ & $\delta$ & $\eta_{\rm dip}$ & $\eta_{\rm prop}$ & $1/f_{\rm B}$ & $\chi^{2}_{\rm red}$ \\
 & & ($\times10^{15}$ G) & (ms) & $\left(\times10^{-2}~M_{\rm \odot}\right)$ & (km) & & & (\%) & (\%) & & \\
\hline
050724 & $10$ & $4.81^{+0.11}_{-0.23}$ & $9.70^{+0.29}_{-0.92}$ & $0.489^{+0.070}_{-0.023}$ & $320^{+5}_{-5}$ & $1.26^{+11.87}_{-1.15}$ & $\left(0.50^{+1.82}_{-0.30}\right)\times10^{-3}$ & $5^{+1}_{-1}$ & $86^{+13}_{-18}$ & $508^{+87}_{-100}$ & $6$ \\ [2pt]
051016B & $100$ & $9.95^{+0.05}_{-0.12}$ & $3.44^{+0.19}_{-0.17}$ & $9.84^{+0.16}_{-0.67}$ & $54^{+4}_{-2}$ & $262.25^{+443.06}_{-244.97}$ & $\left(0.39^{+1.33}_{-0.28}\right)\times10^{-4}$ & $25^{+22}_{-7}$ & $77^{+21}_{-31}$ & $431^{+158}_{-175}$ & $5$ \\ [2pt]
051227 & $100$ & $5.15^{+0.72}_{-0.54}$ & $3.02^{+0.29}_{-0.31}$ & $9.31^{+0.66}_{-1.32}$ & $263^{+10}_{-16}$ & $0.94^{+22.57}_{-0.83}$ & $\left(1.28^{+4.78}_{-0.97}\right)\times10^{-2}$ & $89^{+10}_{-25}$ & $52^{+27}_{-17}$ & $526^{+71}_{-149}$ & $7$ \\ [2pt]
060614 & $1$ & $6.02^{+0.05}_{-0.05}$ & $9.99^{+0.01}_{-0.04}$ & $9.99^{+0.01}_{-0.05}$ & $680^{+6}_{-6}$ & $998.58^{+1.37}_{-6.15}$ & $2.29^{+0.05}_{-0.05}$ & $99^{+1}_{-4}$ & $1$[L] & $322^{+16}_{-8}$ & $19$ \\ [2pt]
061006 & $10$ & $2.60^{+0.53}_{-0.35}$ & $6.83^{+1.88}_{-4.20}$ & $1.14^{+3.82}_{-0.37}$ & $1915^{+82}_{-344}$ & $131.21^{+92.86}_{-51.42}$ & $17.78^{+6.58}_{-4.30}$ & $5$[F] & $40$[F] & $330^{+258}_{-304}$ & $15$ \\ [2pt]
061210 & $10$ & $7.60^{+0.45}_{-0.40}$ & $6.00^{+0.68}_{-0.59}$ & $1.71^{+0.26}_{-0.23}$ & $124^{+3}_{-3}$ & $733.05^{+250.59}_{-285.29}$ & $\left(9.41^{+4.70}_{-3.92}\right)\times10^{-3}$ & $5$[F] & $40$[F] & $241^{+74}_{-55}$ & $23$ \\ [2pt]
070714B & $100$ & $6.58^{+1.56}_{-1.74}$ & $4.91^{+0.80}_{-1.15}$ & $9.32^{+0.65}_{-1.47}$ & $463^{+20}_{-19}$ & $30.93^{+3.52}_{-3.43}$ & $1.68^{+0.12}_{-0.12}$ & $87^{+12}_{-34}$ & $80^{+19}_{-19}$ & $536^{+62}_{-129}$ & $11$ \\ [2pt]
071227 & $1$ & $8.59^{+1.35}_{-2.89}$ & $5.99^{+2.30}_{-3.34}$ & $1.54^{+3.03}_{-0.62}$ & $268^{+22}_{-21}$ & $9.60^{+50.17}_{-9.47}$ & $\left(1.35^{+5.10}_{-0.75}\right)\times10^{-3}$ & $3^{+4}_{-2}$ & $59^{+39}_{-41}$ & $72^{+197}_{-59}$ & $4$ \\ [2pt]
080123 & $100$ & $9.55^{+0.43}_{-1.06}$ & $6.21^{+1.22}_{-1.41}$ & $0.928^{+0.374}_{-0.190}$ & $231^{+7}_{-6}$ & $26.21^{+52.13}_{-24.42}$ & $\left(5.97^{+6.81}_{-2.04}\right)\times10^{-5}$ & $5$[F] & $40$[F] & $158^{+41}_{-60}$ & $6$ \\ [2pt]
080503 & $10$ & $1.97^{+0.45}_{-0.32}$ & $1.85^{+0.59}_{-0.55}$ & $0.33^{+1.08}_{-0.22}$ & $566^{+65}_{-36}$ & $0.42^{+0.21}_{-0.20}$ & $10.35^{+25.85}_{-8.28}$ & $70^{+29}_{-48}$ & $29^{+27}_{-15}$ & $1$[F] & $9$ \\ [4pt]
100212A & $1$ & $9.97^{+0.03}_{-0.13}$ & $7.26^{+0.23}_{-0.23}$ & $9.91^{+0.08}_{-0.33}$ & $163^{+1}_{-1}$ & $978.19^{+21.03}_{-94.38}$ & $\left(7.12^{+0.36}_{-0.68}\right)\times10^{-1}$ & $77^{+22}_{-22}$ & $17^{+5}_{-5}$ & $422^{+166}_{-99}$ & $19$ \\ [2pt]
100522A & $10$ & $9.32^{+0.05}_{-0.09}$ & $9.95^{+0.04}_{-0.19}$ & $0.467^{+0.012}_{-0.003}$ & $75^{+1}_{-1}$ & $5.44^{+3.14}_{-2.50}$ & $\left(8.76^{+2.39}_{-1.20}\right)\times10^{-4}$ & $7^{+3}_{-3}$ & $73^{+26}_{-26}$ & $375^{+205}_{-99}$ & $89$ \\ [2pt]
111121A & $100$ & $4.19^{+0.32}_{-0.28}$ & $4.38^{+0.64}_{-0.46}$ & $8.61^{+1.31}_{-1.44}$ & $812^{+12}_{-12}$ & $99.59^{+30.58}_{-23.43}$ & $\left(2.84^{+0.36}_{-0.33}\right)\times10^{-3}$ & $84^{+16}_{-32}$ & $41^{+12}_{-13}$ & $474^{+120}_{-180}$ & $13$ \\ [2pt]
150424A & $1$ & $9.80^{+0.19}_{-0.73}$ & $5.74^{+0.13}_{-0.41}$ & $9.94^{+0.06}_{-0.22}$ & $826^{+19}_{-22}$ & $339.09^{+61.77}_{-53.67}$ & $6.92^{+0.81}_{-0.78}$ & $19^{+1}_{-2}$ & $99^{+1}_{-4}$ & $594^{+6}_{-22}$ & $11$ \\ [2pt]
160410A & $100$ & $5.02^{+0.81}_{-0.82}$ & $2.33^{+0.78}_{-0.65}$ & $6.23^{+3.36}_{-2.03}$ & $95^{+6}_{-7}$ & $12.25^{+10.49}_{-7.56}$ & $\left(2.20^{+0.63}_{-0.47}\right)\times10^{-2}$ & $24^{+20}_{-15}$ & $78^{+21}_{-39}$ & $419^{+172}_{-223}$ & $6$ \\
\hline
\end{tabular}
\end{table*}

\subsection{GRB 050724}
\label{subsec:050724}

\begin{figure}
\includegraphics[width=\columnwidth]{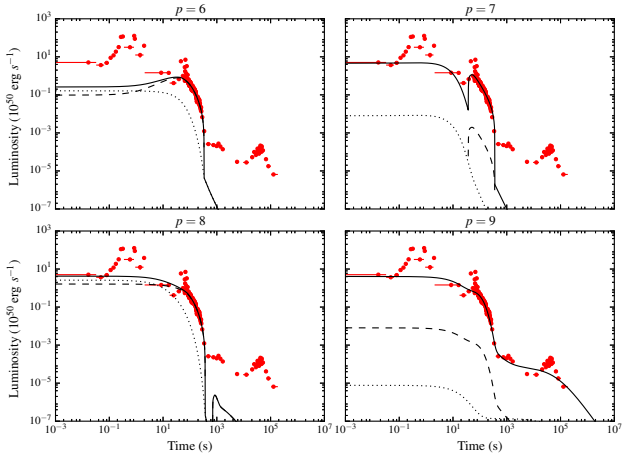}
\caption{Models fitted to GRB 050724 with $n=100$ and $p=6$ (top left), $7$ (top right), $8$ (bottom left), and $9$ (bottom right). Solid line - total luminosity; dashed line - propeller luminosity; dotted line - dipole luminosity; red points - combined BAT and XRT data.}
\label{fig:050724}
\end{figure}

Fig.~\ref{fig:050724} shows a comparison of fits with varying $p$ to GRB 050724 for $n=100$. For $p=6$, the model does a reasonable job of fitting the high luminosity at early times but does not retain enough energy to fit the tail. The fit demanded a large amount of fallback, $\delta=1.65^{+1.26}_{-0.62}$, on a short timescale, $\epsilon=0.13^{+0.04}_{-0.02}$, and a very rapid spin period, $P=0.69$ ms (limit), in order to reach such a high luminosity so soon. Since the fallback mass reaches the disc quickly, there is nothing left in the fallback budget to provide energy for the late-time emission. $p=7$ and $8$ provide improved fits to the early-time luminosity but again fail to fit the fading tail despite the additional parameters being pushed to the higher end of their limits, e.g.~$1/f_{\rm B}=578^{+21}_{-71}$ for $p=7$ and $\eta_{\rm prop}=99^{+1}_{-5}$\% for $p=8$ $\left(\eta_{\rm dip}=12^{+2}_{-2}\%\right)$. $p=9$ is the only model that succeeds in fitting the tail but still requires a highly efficient emission mechanism for the propeller, $\eta_{\rm prop}=86^{+13}_{-18}$\%, and a very narrow beaming angle, $1/f_{\rm B}=502^{+93}_{-103}$.

It is interesting to note the late-time giant flare within the tail of GRB 050724 that the model has not been able to fit. At present, the phenomena that cause such large outbursts at these late times are still poorly understood (see \citealt{falcone06}, \citealt{curran08}, and \citealt{chincarini10}).

\subsection{GRB 060614}
\label{subsec:060614}

GRB 060614 poses a challenge to typical long/short classification scheme since it has a duration of $\sim100$ s but the hard spectrum and lack of supernova connection are more indicative of the short classification \citep{mangano07,zhang07b,xu09}.

\begin{figure}
\includegraphics[width=\columnwidth]{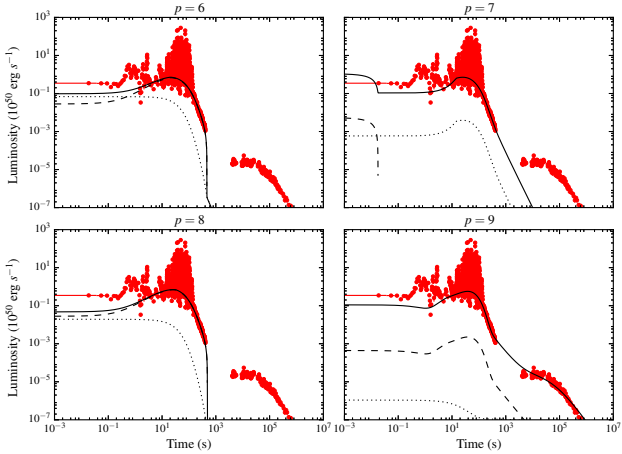}
\caption{Models fitted to GRB 060614 with $n=100$ and $p=6$ (top left), $7$ (top right), $8$ (bottom left), and $9$ (bottom right). Solid line - total luminosity; dashed line - propeller luminosity; dotted line - dipole luminosity; red points - combined BAT and XRT data.}
\label{fig:060614}
\end{figure}

Fig.~\ref{fig:060614} presents model fits of varying $p$ and $n=100$ to data for GRB 060614. $p=6$ provides a good fit to the early-time luminosity but after $\sim 100$ s, its energy reservoir is depleted and the light curve rapidly drops off before fitting the tail. This demands a rapid spin period, $P=0.90^{+0.01}_{-0.01}$ ms, and a large amount of fallback mass, $\delta=49.61^{+0.38}_{-1.63}$, reaching the disc on a short timescale, $\epsilon=0.31^{+0.03}_{-0.03}$. $p=7$ adds more structure to the early-time luminosity and has a more gradual decrease of emission but still fails to reach the tail, whereas, $p=8$ is very much a repeat of $p=6$ and offers no improvement. Again, $p=9$ offers the best results for fitting to the tail but requires a very efficient emission mechanism for the propeller, $\eta_{\rm prop}=100\%$ (limit), and a moderate beaming fraction, $1/f_{\rm B}=251^{+6}_{-7}$. Oddly, this model requires the least efficient dipole emission as well, $\eta_{\rm dip}=1\%$ (limit). This is probably due to the difference in EE and dipole luminosity being the greatest in GRB 060614 and so the model has to do something to achieve a drop in luminosity spanning several orders of magnitude while maintaining parameters that can produce bright, early emission.

GRB 060614 continues to be a very odd case when we examine its best fit parameters in Table~\ref{tab:results_pars} relating to the fit in Fig.~\ref{fig:results} and $n=10$. It is one of the slowest rotating candidates with one of the most massive and most slowly fed discs. Also, the dipole and propeller emission efficiencies have completely reversed roles with $\eta_{\rm dip}=99^{+1}_{-4}\%$ and $\eta_{\rm prop}=1\%$ (limit). The propeller's main job is to modulate the spin in order to achieve the desired luminosities. Since the propeller plays no role in this particular fit, this indicates that that has been completely taken over by the fallback.

\subsection{GRB 111121A}
\label{subsec:111121A}

\begin{figure}
\includegraphics[width=\columnwidth]{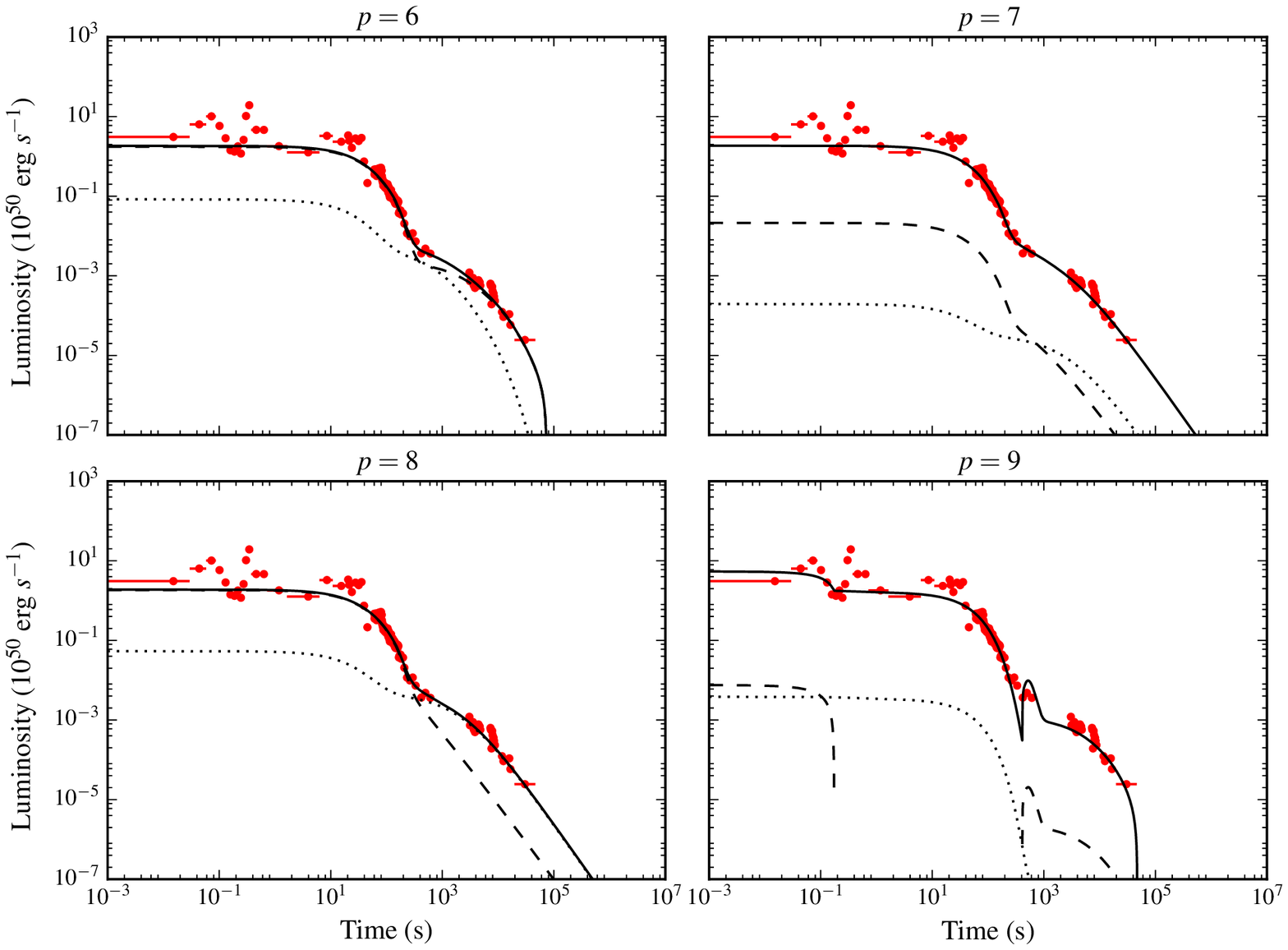}
\caption{Models fitted to GRB 111121A with $n=100$ and $p=6$ (top left), $7$ (top right), $8$ (bottom left), and $9$ (bottom right). Solid line - total luminosity; dashed line - propeller luminosity; dotted line - dipole luminosity; red points - combined BAT and XRT data.}
\label{fig:111121A}
\end{figure}

Fig.~\ref{fig:111121A} presents model fits of varying $p$ and $n=100$ to data for GRB 111121A. This is an example of the model behaving well across all values of $p$. Despite the fits for $p=6,~7$ and $8$ looking very similar, the parameters derived from the fits vary quite significantly. For $p=6$ and $8$, relatively small values of magnetic field are recovered, $B=\left(2.00^{+0.12}_{-0.11}\right)\times10^{15}$ G and $B=\left(1.53^{+1.40}_{-0.31}\right)\times10^{15}$ G respectively, whereas $p=7$ has a large magnetic field of $B=\left(8.15^{+1.77}_{-4.65}\right)\times10^{15}$ G. The initial spin values for these fits also follow a similar pattern with spins near the break-up limit for $p=6$ and $8$, $P_{\rm i}=0.69$ ms (limit) and $P_{\rm i}=0.89^{+0.16}_{-0.17}$ ms respectively, and a much slower spin for $p=7$, $P_{\rm i}=6.36^{+1.96}_{-3.76}$ ms.

Lastly, the $p=9$ fit has derived parameters in the moderate region of parameter space, $B=\left(4.19^{+0.32}_{-0.28}\right)\times10^{15}$ G and $P_{\rm i}=4.38^{+0.64}_{-0.46}$ ms. It has a slowly fed disc with a small amount of fallback mass, $\epsilon=99.59^{+30.58}_{-23.43}$ and $\delta=\left(2.84^{+0.36}_{-0.33}\right)\times10^{-3}$. We derive a propeller efficiency consistent with the value used to \citet{gompertz14} of $\eta_{\rm prop}=41.48^{+11.99}_{-13.27}\%$ but the fit requires a much higher dipole efficiency of $\eta_{\rm dip}=83.60^{+15.65}_{-31.94}\%$ and a narrow jet opening angle of $1/f_{\rm B}=474^{+120}_{-180}$. However, this fit has introduced a flare at roughly the $1000$ s mark which could be indicative of over-fitting.

\subsection{Refitting excluding early-time data}
\label{subsec:refit}

The results presented in Table~\ref{tab:results_pars} are consistently pushing the upper bounds for the initial disc mass, $M_{\rm D,i}$. This is most likely due to the model's need to have a high accretion rate at early-times in order to reach the high luminosities at those times. Since the emission produced at these times is usually attributed to internal shocks and energy drawn from the merger rather than magnetic particle acceleration, fitting these high early-time luminosities may not strictly be within the remit of the model. We therefore chose to refit the sample excluding some of the early-time data.

We chose an arbitrary cut-off of $10$ seconds to define the on-set of EE after the prompt emission. This meant we avoided making an arbitrary cut for each individual burst since EE isn't currently well defined. The fits were performed for $p=6,~7,~8$ and $9$ and $n=1$ for comparison with the work in \citet{gompertz14}.

\begin{table}
\centering
\caption{AICc values for fits to the SGRBEE sample with varying $p$ values and $n=1$ with data $<10$s excluded. Values in bold face are the minimum value for each GRB. $^*$GRB 061210 has fewer data points than free parameters resulting in a negative AICc value which was not considered when choosing the best fit.}
\begin{tabular}{lrrrr}
\hline
GRB & $p=6$ & $p=7$ & $p=8$ & $p=9$ \\
\hline
050724 & $1,611$ & $1,489$ & $1,561$ & \boldmath$1,259$ \\
051016B & $340$ & $252$ & $310$ & \boldmath$153$ \\
051227 & $178$ & \boldmath$53$ & $59$ & $70$ \\
060614 & $48,086$ & $43,738$ & $43,728$ & \boldmath$43,610$ \\
061006 & $177$ & \boldmath$90$ & $114$ & $123$ \\
061210 & $-17^*$ & $181$ & \boldmath$66$ & $357$ \\
070714B & $203$ & $215$ & \boldmath$177$ & $195$ \\
071227 & $112$ & \boldmath$89$ & $100$ & $101$ \\
080123 & $354$ & $308$ & \boldmath$298$ & $319$ \\
080503 & $2,281$ & $2,375$ & \boldmath$2,157$ & $2,339$ \\
100212A & $8,198$ & $7,602$ & $7,771$ & \boldmath$7,073$ \\
100522A & $8,530$ & $7,401$ & $7,725$ & \boldmath$6,322$ \\
111121A & $872$ & \boldmath$782$ & $819$ & $787$ \\
150424A & $366$ & $279$ & $354$ & \boldmath$251$ \\
160410A & $495$ & \boldmath$149$ & $212$ & $222$ \\
\hline
\end{tabular}
\label{tab:10s}
\end{table}

\begin{figure*}
\includegraphics[width=\textwidth]{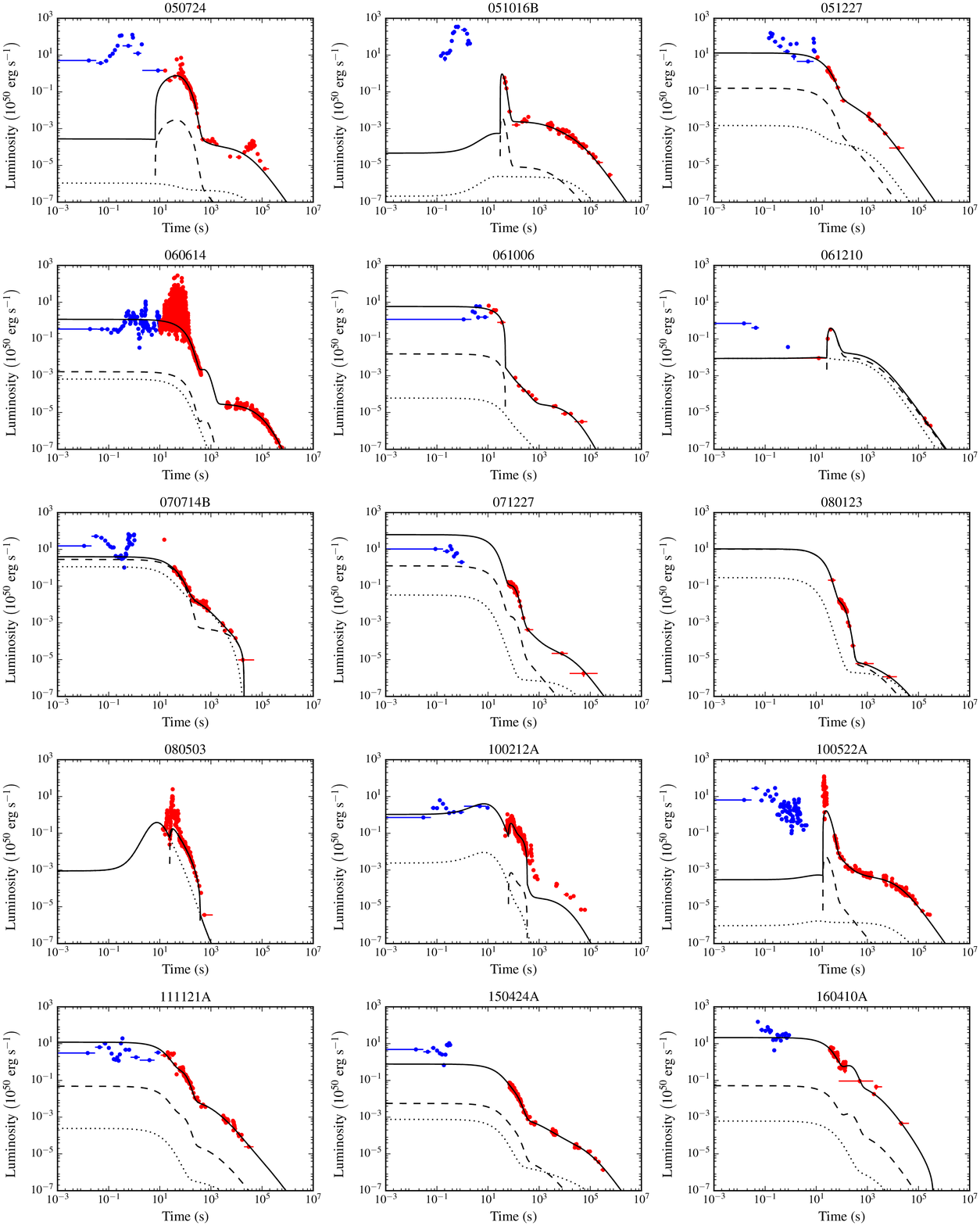}
\caption{Global best fit models produced from fitting to the SGRBEE sample for $n=1$ and excluding data $<10$ s (bold values in Table~\ref{tab:10s}). Solid, black line - total luminosity; dashed, black line - propeller luminosity; dotted, black line - dipole luminosity. Points are combined BAT and XRT data: red points have been included in the fitting, blue points were excluded.}
\label{fig:10s}
\end{figure*}

Table~\ref{tab:10s} presents the AICc values of the refits. The best fits (bold values) from Table~\ref{tab:10s} are plotted in Fig.~\ref{fig:10s} and the parameters derived from these fits are presented in Table~\ref{tab:results_10s} with the $\chi^{2}_{\rm red}$ goodness of fit statistic. GRB 061210 has very few data points and excluding data $<10$ seconds means that there are fewer data points than free parameters which resulted in negative AICc and $\chi^{2}_{\rm red}$ values. Therefore, it is shown here for consistency rather than as a statistically significant result.

\begin{table*}
\centering
\caption{Parameters derived from the best fitting models to the SGRBEE sample for $n=1$ and excluding data $<10$s. Uncertainties represent a 95\% confidence interval. Values marked with an [L] are a parameter limit; those marked with an [F] were fixed during fitting. $\chi^{2}_{\rm red}$ values are shown to indicate goodness of fit. $^*$GRB 061210 has fewer data points than free parameters resulting in a negative $\chi^{2}_{\rm red}$ value.}
\begin{tabular}{lcccccccccc}
\hline
GRB & $B$ & $P_{\rm i}$ & $M_{\rm D,i}$ & $R_{\rm D}$ & $\epsilon$ & $\delta$ & $\eta_{\rm dip}$ & $\eta_{\rm prop}$ & $1/f_{\rm B}$ & $\chi^{2}_{\rm red}$ \\
 & ($\times10^{-15}$ G) & (ms) & $\left(\times10^{-2}~M_{\odot}\right)$ & (km) & & & (\%) & (\%) & \\
\hline
050724 & $3.30^{+0.48}_{-0.36}$ & $9.94^{+0.06}_{-0.26}$ & $0.167^{+0.042}_{-0.028}$ & $383^{+7}_{-7}$ & $0.94^{+12.75}_{-0.83}$ & $\left(0.85^{+3.15}_{-0.73}\right)\times10^{-3}$ & $1$[L] & $98^{+2}_{-8}$ & $253^{+35}_{-22}$ & $5$ \\ [2pt]
051016B & $1.18^{+0.68}_{-0.48}$ & $9.09^{+0.87}_{-2.84}$ & $2.47^{+4.73}_{-1.63}$ & $59^{+4}_{-3}$ & $635.72^{+234.17}_{-216.98}$ & $\left(1.00^{+0.91}_{-0.51}\right)\times10^{-3}$ & $1$[L] & $94^{+6}_{-20}$ & $220^{+296}_{-144}$ & $2$ \\ [2pt]
051227 & $8.86^{+1.10}_{-3.19}$ & $4.00^{+1.66}_{-1.84}$ & $0.67^{+1.84}_{-0.31}$ & $134^{+35}_{-26}$ & $4.98^{+456.99}_{-4.86}$ & $\left(0.30^{+1.31}_{-0.30}\right)\times10^{-1}$ & $5$[F] & $40$[F] & $82^{+77}_{-57}$ & $2$ \\ [2pt]
060614 & $6.21^{+0.41}_{-1.03}$ & $8.30^{+1.14}_{-2.48}$ & $0.755^{+0.453}_{-0.117}$ & $1680^{+23}_{-23}$ & $562.87^{+161.29}_{-103.91}$ & $\left(1.00^{+0.14}_{-0.10}\right)\times10^{-2}$ & $84^{+15}_{-52}$ & $8^{+2}_{-4}$ & $506^{+91}_{-225}$ & $20$ \\ [2pt]
061006 & $7.16^{+0.81}_{-1.07}$ & $7.97^{+1.63}_{-3.97}$ & $2.07^{+3.29}_{-0.56}$ & $1908^{+89}_{-372}$ & $136.05^{+94.65}_{-54.63}$ & $20.98^{+7.99}_{-5.34}$ & $5$[F] & $40$[F] & $380^{+209}_{-311}$ & $8$ \\ [2pt]
061210 & $0.75^{+0.27}_{-0.24}$ & $0.80^{+0.26}_{-0.11}$ & $7.04^{+2.76}_{-2.20}$ & $128^{+143}_{-42}$ & $182.82^{+748.09}_{-165.31}$ & $\left(1.47^{+6.69}_{-1.13}\right)\times10^{-1}$ & $7^{+3}_{-3}$ & $91^{+9}_{-25}$ & $1$[F] & $-33^*$ \\ [2pt]
070714B & $4.97^{+1.20}_{-1.45}$ & $1.00^{+0.15}_{-0.27}$ & $4.24^{+2.93}_{-0.90}$ & $320^{+29}_{-25}$ & $520.96^{+453.38}_{-494.48}$ & $\left(1.55^{+0.92}_{-0.65}\right)\times10^{-1}$ & $48^{+42}_{-31}$ & $80^{+19}_{-36}$ & $1$[F] & $2$ \\ [2pt]
071227 & $8.40^{+1.52}_{-2.37}$ & $1.79^{+1.99}_{-0.80}$ & $4.98^{+4.42}_{-2.80}$ & $250^{+25}_{-23}$ & $1.78^{+28.53}_{-1.67}$ & $\left(0.95^{+4.28}_{-0.61}\right)\times10^{-3}$ & $5$[F] & $40$[F] & $49^{+51}_{-33}$ & $2$ \\ [2pt]
080123 & $7.08^{+0.18}_{-0.25}$ & $0.91^{+0.07}_{-0.04}$ & $9.84^{+0.16}_{-0.61}$ & $254^{+5}_{-4}$ & $62.03^{+117.14}_{-58.06}$ & $\left(1.56^{+1.90}_{-1.19}\right)\times10^{-4}$ & $4^{+3}_{-3}$ & $98^{+2}_{-9}$ & $1$[F] & $6$ \\ [2pt]
080503 & $5.45^{+0.55}_{-1.29}$ & $6.95^{+2.91}_{-4.62}$ & $6.27^{+3.61}_{-5.58}$ & $59^{+2}_{-5}$ & $0.108^{+0.079}_{-0.008}$ & $4.21^{+37.64}_{-2.15}$ & $73^{+26}_{-54}$ & $60^{+20}_{-36}$ & $1$[F] & $9$ \\ [2pt]
100212A & $9.98^{+0.02}_{-0.07}$ & $7.50^{+2.39}_{-4.32}$ & $9.96^{+0.04}_{-0.17}$ & $163^{+1}_{-1}$ & $980.05^{+19.22}_{-87.98}$ & $\left(7.07^{+0.23}_{-0.62}\right)\times10^{-1}$ & $79^{+20}_{-20}$ & $17^{+5}_{-4}$ & $441^{+148}_{-94}$ & $19$ \\ [2pt]
100522A & $3.06^{+0.09}_{-0.07}$ & $9.93^{+0.07}_{-0.33}$ & $0.509^{+0.076}_{-0.105}$ & $63^{+2}_{-1}$ & $0.22^{+1.21}_{-0.11}$ & $\left(3.11^{+2.10}_{-1.87}\right)\times10^{-3}$ & $1$[L] & $99^{+1}_{-3}$ & $316^{+26}_{-19}$ & $37$ \\ [2pt]
111121A & $9.03^{+0.92}_{-2.25}$ & $6.34^{+1.64}_{-2.71}$ & $1.14^{+1.22}_{-0.31}$ & $292^{+16}_{-14}$ & $32.44^{+7.02}_{-5.21}$ & $\left(1.77^{+0.36}_{-0.26}\right)\times10^{-2}$ & $5$[F] & $40$[F] & $247^{+119}_{-167}$ & $7$ \\ [2pt]
150424A & $9.19^{+0.68}_{-1.03}$ & $8.99^{+0.97}_{-2.76}$ & $0.544^{+0.421}_{-0.115}$ & $434^{+61}_{-43}$ & $20.40^{+9.75}_{-7.56}$ & $\left(1.88^{+0.66}_{-0.59}\right)\times10^{-2}$ & $61^{+37}_{-45}$ & $13^{+14}_{-9}$ & $122^{+331}_{-69}$ & $2$ \\ [2pt]
160410A & $3.58^{+0.63}_{-0.95}$ & $3.17^{+0.80}_{-1.43}$ & $3.32^{+3.94}_{-0.89}$ & $826^{+79}_{-87}$ & $21.83^{+199.41}_{-19.02}$ & $\left(3.65^{+5.63}_{-2.19}\right)\times10^{-2}$ & $5$[F] & $40$[F] & $410^{+181}_{-285}$ & $3$ \\ [2pt]
\hline
\end{tabular}
\label{tab:results_10s}
\end{table*}

As is shown in Fig.~\ref{fig:10s}, the result of excluding the early-time data is to produce more light curves of the \emph{humped} morphology than the \emph{sloped} or \emph{classic} variety in Fig.~\ref{fig:results}. But most surprisingly, this experiment did not succeed in reducing $M_{\rm D,i}$ as expected, suggesting the extra mass is a result of another change in the model, most likely the use of Equation~(\ref{eq:ndip}) instead of Equation~(\ref{eq:bucc}). Equation~(\ref{eq:bucc}) enhances the dipole spin-down and mass-loss resulting in a lower initial disc mass.

\subsection{Refitting with enhanced dipole torque}
\label{subsec:bucc}

\begin{figure*}
\includegraphics[width=\textwidth]{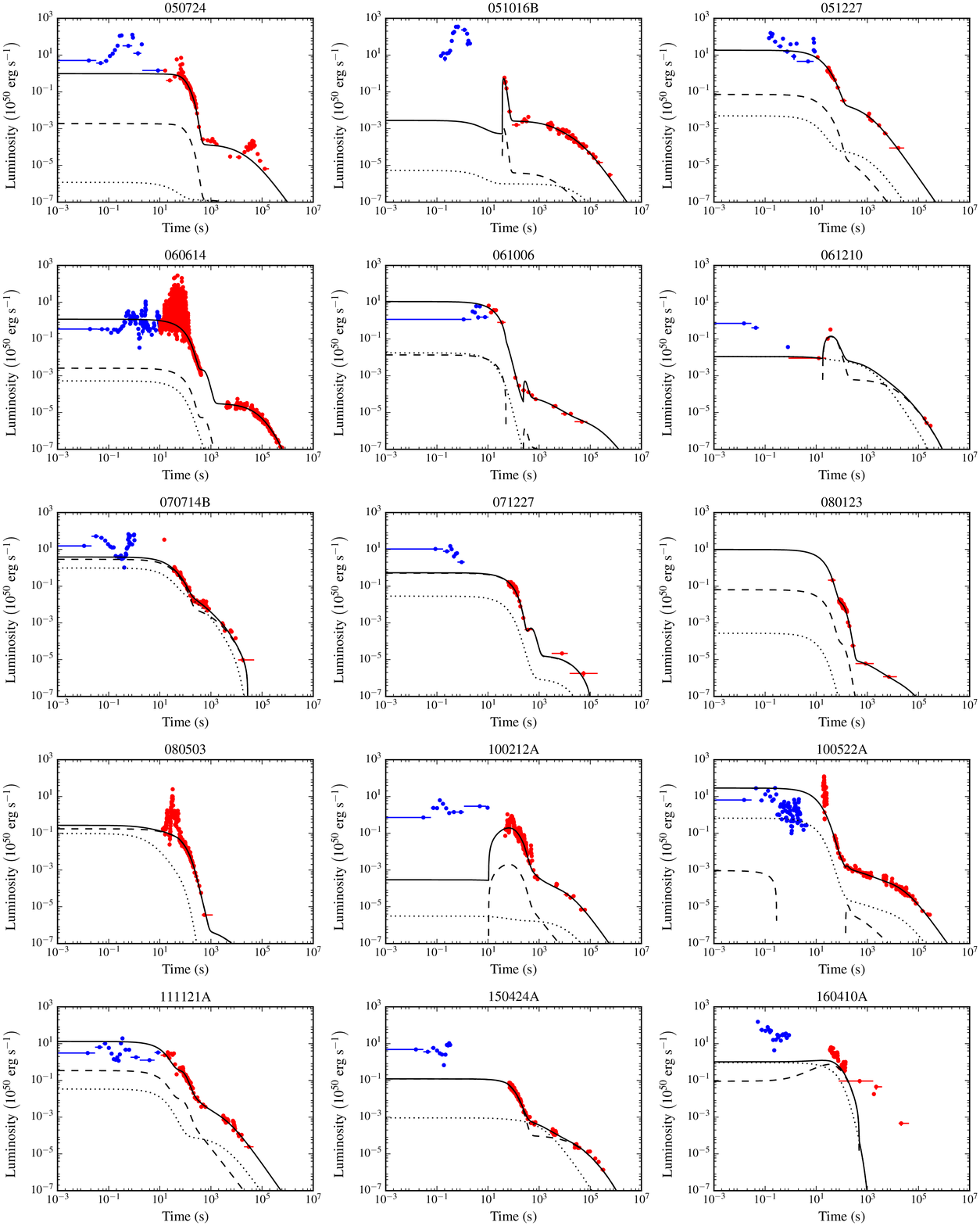}
\caption{Results of fitting to the SGRBEE sample for the best global fits (bold values in Table~\ref{tab:bucc_aicc}) excluding data $<10$s and using Equation~(\ref{eq:bucc}) for the dipole torque. Solid, black line - total luminosity; dashed, black line - propeller luminosity; dotted, black line - dipole luminosity. Data points are combined BAT and XRT data: blue points have been excluded from the fit, red points were included.}
\label{fig:bucc_results}
\end{figure*}

For direct comparison with \citet{gompertz14}, the sample was fitted once more using the enhanced dipole torque in Equation~(\ref{eq:bucc}) from \citet{bucciantini06} for $n=1$ and $p=6,~7,~8$ and $9$. The AICc values for the fits are presented in Table~\ref{tab:bucc_aicc}, the best fits from this table are shown in Fig.~\ref{fig:bucc_results}, and the parameters derived from those fits are presented in Table~\ref{tab:bucc_pars}.

\begin{table}
\centering
\caption{AICc values for fits to the SGRBEE sample excluding data $<10$ s and using Equation~(\ref{eq:bucc}) for the dipole torque. $^*$GRB 061210 has fewer data points than free parameters and so these statistics should be treat with caution.}
\begin{tabular}{lrrrr}
\hline
GRB & $p=6$ & $p=7$ & $p=8$ & $p=9$ \\
\hline
050724 & $5,927$ & $17,516$ & $1,636$ & \boldmath$1,287$ \\
051016B & $396$ & $2,327$ & $444$ & \boldmath$147$ \\
051227 & $493$ & $1,000$ & $192$ & \boldmath$64$ \\
060614 & $50,687$ & $88,449$ & $44,055$ & \boldmath$43,667$ \\
061006 & $294$ & $854$ & $424$ & \boldmath$111$ \\
061210$^*$ & $372$ & $1,236$ & \boldmath$96$ & $1,532$ \\
070714B & $880$ & $3,622$ & \boldmath$212$ & $1,402$ \\
071227 & \boldmath$115$ & $1,293$ & $161$ & $202$ \\
080123 & $355$ & $6,162$ & $752$ & \boldmath$290$ \\
080503 & $3,186$ & $13,299$ & \boldmath$2,336$ & $3,670$ \\
100212A & $9,035$ & $35,074$ & $8,757$ & \boldmath$8,037$ \\
100522A & $8,709$ & $22,489$ & $8,808$ & \boldmath$6,391$ \\
111121A & $2,498$ & $8,441$ & $898$ & \boldmath$757$ \\
150424A & $398$ & $5,861$ & \boldmath$252$ & $267$ \\
160410A & $1,198$ & $1,479$ & \boldmath$717$ & $971$ \\
\hline
\end{tabular}
\label{tab:bucc_aicc}
\end{table}

\begin{table*}
\centering
\caption{Parameters derived from fits to SGRBEE sample for the globals fits (bold values in Table~\ref{tab:bucc_aicc}) excluding data $<10$s and using Equation~(\ref{eq:bucc}) for the dipole torque. Uncertainties are a $95\%$ confidence interval and values marked with an [L] are a parameter limit. $^*$GRB 061210 has fewer data points than parameters resulting in a negative $\chi^{2}_{\rm red}$ value.}
\label{tab:bucc_pars}
\begin{tabular}{lcccccccccr}
\hline
GRB & $B$ & $P_{\rm i}$ & $M_{\rm D,i}$ & $R_{\rm D}$ & $\epsilon$ & $\delta$ & $\eta_{\rm dip}$ & $\eta_{\rm prop}$ & $1/f_{\rm B}$ & $\chi^{2}_{\rm red}$ \\
& ($\times10^{15}$ G) & (ms) & $\left(\times10^{-2}~M_{\odot}\right)$ & (km) & & & (\%) & (\%) & & \\
\hline
050724 & $2.75^{+0.26}_{-0.31}$ & $8.91^{+0.63}_{-1.12}$ & $0.100^{+0.021}_{-0.005}$ & $378^{+8}_{-8}$ & $983.76^{+15.64}_{-67.62}$ & $\left(5.70^{+1.15}_{-1.13}\right)\times10^{-3}$ & $1$[L] & $97^{+3}_{-10}$ & $522^{+73}_{-128}$ & $5$ \\ [2pt]
051016B & $0.36^{+0.20}_{-0.09}$ & $2.33^{+0.61}_{-0.57}$ & $2.62^{+2.35}_{-1.56}$ & $52^{+5}_{-2}$ & $772.31^{+205.95}_{-259.35}$ & $\left(4.15^{+5.45}_{-1.97}\right)\times10^{-4}$ & $1$[L] & $92^{+7}_{-23}$ & $524^{+73}_{-213}$ & $2$ \\ [2pt]
051227 & $8.12^{+1.80}_{-3.86}$ & $4.91^{+3.23}_{-3.01}$ & $0.26^{+0.87}_{-0.15}$ & $159^{+60}_{-40}$ & $13.32^{+732.02}_{-13.21}$ & $\left(0.77^{+6.62}_{-0.77}\right)\times10^{-2}$ & $45^{+51}_{-36}$ & $53^{+44}_{-44}$ & $238^{+327}_{-187}$ & $2$ \\ [2pt]
060614 & $3.03^{+0.36}_{-0.39}$ & $6.05^{+1.60}_{-1.42}$ & $0.31^{+0.13}_{-0.08}$ & $1297^{+17}_{-17}$ & $632.65^{+119.29}_{-105.42}$ & $\left(1.53^{+0.14}_{-0.14}\right)\times10^{-2}$ & $80^{+19}_{-41}$ & $19^{+10}_{-8}$ & $383^{+208}_{-222}$ & $20$ \\ [2pt]
061006 & $3.57^{+0.61}_{-0.75}$ & $2.67^{+0.93}_{-1.02}$ & $1.08^{+0.96}_{-0.36}$ & $423^{+15}_{-14}$ & $52.71^{+51.20}_{-29.47}$ & $\left(2.38^{+0.65}_{-0.61}\right)\times10^{-3}$ & $73^{+26}_{-48}$ & $6^{+6}_{-3}$ & $347^{+240}_{-273}$ & $11$ \\ [2pt]
061210 & $0.40^{+0.03}_{-0.03}$ & $0.69$[L] & $1.19^{+0.25}_{-0.22}$ & $211^{+89}_{-59}$ & $815.21^{+177.54}_{-528.85}$ & $\left(3.21^{+3.07}_{-0.88}\right)\times10^{-1}$ & $17^{+3}_{-3}$ & $99^{+1}_{-3}$ & $1$[F] & $-43^*$ \\ [2pt]
070714B & $1.77^{+0.17}_{-0.15}$ & $0.69$[L] & $3.43^{+0.38}_{-0.82}$ & $307^{+26}_{-22}$ & $41.83^{+598.49}_{-21.04}$ & $\left(1.21^{+0.92}_{-0.32}\right)\times10^{-1}$ & $75^{+24}_{-59}$ & $99^{+1}_{-5}$ & $1$[F] & $2$ \\ [2pt]
071227 & $1.19^{+0.27}_{-0.13}$ & $0.70^{+0.04}_{-0.01}$ & $5.46^{+0.20}_{-0.60}$ & $1144^{+97}_{-116}$ & $250.23^{+674.92}_{-225.49}$ & $\left(0.20^{+1.25}_{-0.13}\right)\times10^{-1}$ & $5$[F] & $40$[F] & $1$[F] & $3$ \\ [2pt]
080123 & $8.31^{+1.61}_{-2.20}$ & $5.67^{+2.71}_{-2.52}$ & $0.32^{+0.38}_{-0.13}$ & $244^{+7}_{-6}$ & $54.14^{+91.66}_{-48.87}$ & $\left(1.26^{+1.09}_{-0.60}\right)\times10^{-4}$ & $4^{+7}_{-3}$ & $66^{+32}_{-40}$ & $154^{+351}_{-112}$ & $6$ \\ [2pt]
080503 & $9.10^{+0.87}_{-4.06}$ & $1.06^{+0.74}_{-0.35}$ & $0.75^{+0.92}_{-0.12}$ & $767^{+23}_{-40}$ & $36.89^{+828.92}_{-36.76}$ & $\left(0.01^{+16.16}_{-0.01}\right)\times10^{-1}$ & $2^{+3}_{-1}$ & $68^{+21}_{-43}$ & $1$[F] & $9$ \\ [2pt]
100212A & $0.73^{+0.08}_{-0.07}$ & $3.91^{+0.37}_{-0.34}$ & $0.11^{+0.02}_{-0.01}$ & $550^{+15}_{-15}$ & $0.37^{+0.95}_{-0.25}$ & $\left(2.46^{+3.05}_{-1.34}\right)\times10^{-2}$ & $1$[L] & $77^{+21}_{-26}$ & $94^{+50}_{-26}$ & $22$ \\ [2pt]
100522A & $1.53^{+0.29}_{-0.12}$ & $0.74^{+0.15}_{-0.05}$ & $8.45^{+1.22}_{-2.38}$ & $243^{+6}_{-5}$ & $18.73^{+10.36}_{-6.57}$ & $\left(4.72^{+0.56}_{-0.50}\right)\times10^{-3}$ & $89^{+11}_{-24}$ & $1$[L] & $43^{+22}_{-10}$ & $37$ \\ [2pt]
111121A & $2.21^{+0.26}_{-0.22}$ & $1.45^{+0.34}_{-0.26}$ & $1.97^{+0.64}_{-0.51}$ & $247^{+15}_{-12}$ & $0.20^{+0.73}_{-0.10}$ & $\left(5.00^{+3.77}_{-3.18}\right)\times10^{-2}$ & $32^{+22}_{-24}$ & $60^{+38}_{-44}$ & $35^{+91}_{-18}$ & $6$ \\ [2pt]
150424A & $0.38^{+0.08}_{-0.07}$ & $0.75^{+0.12}_{-0.05}$ & $0.39^{+0.11}_{-0.04}$ & $540^{+22}_{-21}$ & $951.95^{+46.20}_{-168.74}$ & $\left(5.13^{+0.88}_{-0.84}\right)\times10^{-1}$ & $2^{+1}_{-1}$ & $91^{+9}_{-21}$ & $1$[F] & $2$ \\ [2pt]
160410A & $1.49^{+0.14}_{-0.13}$ & $0.69$[L] & $0.17^{+0.53}_{-0.06}$ & $544^{+563}_{-322}$ & $0.57^{+6.17}_{-0.46}$ & $20.42^{+22.88}_{-16.15}$ & $99^{+1}_{-4}$ & $100$[L] & $1$[F] & $17$ \\
\hline
\end{tabular}
\end{table*}

Including Equation~(\ref{eq:bucc}) in the model provides a marginal improvement in fitting, e.g.~the tail of GRB 100212A is matched more closely in Fig.~\ref{fig:bucc_results} than Fig.~\ref{fig:10s}, though in some cases it performs much worse, e.g.~GRB 160410A. The initial disc mass $M_{\rm D,i}$ is reduced by approximately an order of magnitude across the sample. This is a reflection of the enhanced energy output facilitated by Equation~(\ref{eq:bucc}). Equation~(\ref{eq:bucc}) does not produce a dramatic change in the morphology or energetics of the fits, nor does it significantly improve the fit statistics. However, the derived disc masses are more broadly in line with previous work (e.g.~\citealt{rosswog07}).

\subsection{The B-P landscape}
\label{subsec:BP}

\begin{figure}
\includegraphics[width=\columnwidth]{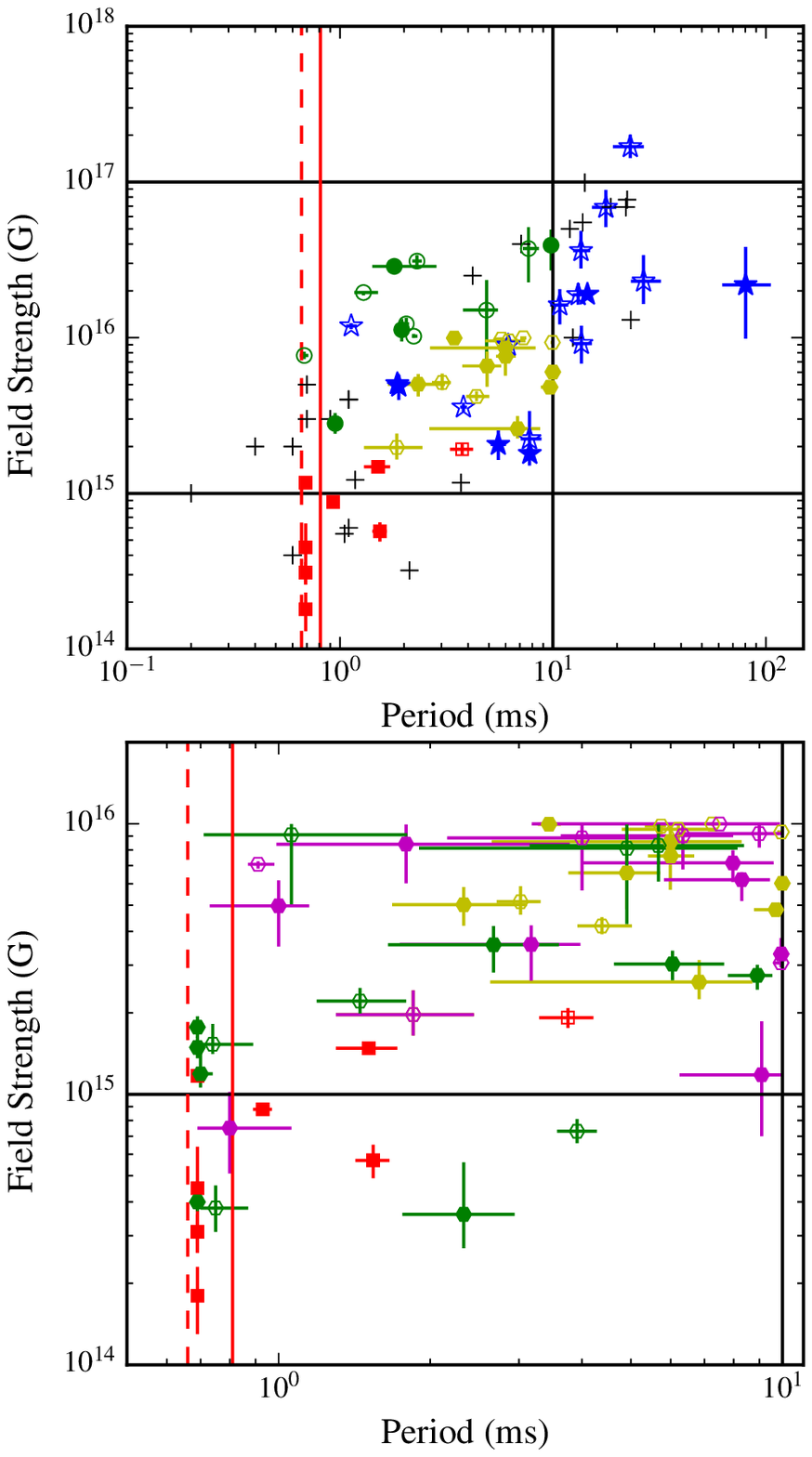}
\caption{Plots of magnetic field strength versus initial spin period. The solid (dashed) red line represents the break-up period for a collapsar (binary merger) progenitor \citet{lattimer04}. \textbf{Top panel} - blue stars: stable magnetars and green circles: unstable magnetars which collapse to form a black hole \citet{rowlinson13}. Black `+' symbols are the LGRB candidates identified by \citet{lyons10,dallosso11}. Red squares (both panels) show the values found in \citet{gompertz14}. Yellow hexagons (both panels) represent the magnetic fields and initial spin periods of this work for the global best fit values in Table~\ref{tab:results_pars}. \textbf{Bottom panel} - magenta hexagons are the $B$ and $P_{\rm i}$ values for fits excluding data $<10$ s in Table~\ref{tab:results_10s}; green hexagons are $B$ and $P_{\rm i}$ values for fits excluding data $<10$ s and including Equation~(\ref{eq:bucc}) in Table~\ref{tab:bucc_pars}. Filled symbols have observed redshifts, open symbols use the sample average redshift, which is $z=0.39$ for extended bursts and $z=0.72$ for the short bursts from \citet{rowlinson13}.}
\label{fig:PvsB}
\end{figure}

Fig.~\ref{fig:PvsB} shows where the results of this work fall in relation to other GRBs in both the long and short classifications. It needs to be noted that the results from \citet{gompertz14} used fixed efficiencies of $\eta_{\rm dip} = 5\%$ and $\eta_{\rm prop}=40\%$, whereas the work done in \citet{rowlinson13} uses 100\% efficiency instead, and our efficiencies have been free parameters in most fitting procedures. Also, \citet{gompertz14} used Equation~(\ref{eq:bucc}) which enhances the dipole spin-down and so these results appear to occupy their own region of low magnetic field and spin period. Hence, conclusions drawn from this plot require some caution.

However, Fig.~\ref{fig:PvsB} does show us that our results occupy a region of moderate to high magnetic field and spin period, indicating that the fallback accretion relaxes the constraints on the initial spin of the magnetar (i.e.~it does not need to be born near the break-up period) since it will be spun-up by the fallback regardless. Though this result could be due to either the addition of a $t^{-5/3}$ fallback accretion profile or our inclusion of beaming as a fitting parameter. The results of this work still do not approach the same same region as \citet{gompertz14} even when early-time, high luminosity data is excluded and Equation~(\ref{eq:bucc}) is used which consolidates that the shift in $B$-$P$ parameter space is due to the inclusion of fallback accretion.

\section{Conclusions}
\label{sec:concs}

We have modified the magnetar propeller model to include fallback accretion, examined the effect these changes have on model light curves and used a MCMC to fit the model to a sample of short GRBs exhibiting extended emission for a range of free parameters and ``sharpness'' of propeller. We have found that the parameters derived from the fits produced by the propeller model with fallback accretion are consistent with theoretical predictions for magnetars.

Our model can cope with long, dipole plateaux and flare-like variability but struggles with the early-time, short-timescale variability. However, since this variability is usually present in the prompt emission which is generally attributed to internal shocks rather than magnetic acceleration of particles, it is not strictly within the remit of the model to fit it.

The addition of fallback accretion provides a noticeable improvement in matching light curves compared to those presented in \citet{gompertz14} and fallback accretion may play a pivotal role in explaining the features of extended emission light curves. Our model uses a smoothed representation of fallback disc feeding as a simplest case scenario. A more ``clumpy'' representation could potentially be more physical and useful to explain phenomena such as flares \citep{dallosso17}.

\section*{Acknowledgements}

The authors would like to thank the reviewers for their helpful and constructive comments. SLG would like to thank Dr. Mark Wilkinson at the University of Leicester for many instructive conversations and acknowledge funding from the Weizmann Institute and the University of Leicester. PTO would like to acknowledge funding from STFC. This research used the ALICE High Performance Computing Facility at the University of Leicester. The work makes use of data supplied by the UKSSDC at the University of Leicester and the \emph{Swift} satellite. \emph{Swift}, launched in November 2004, is a NASA mission in partnership with the Italian Space Agency and the UK Space Agency. \emph{Swift} is managed by NASA Goddard. Penn State University controls science and flight operations from the Mission Operations Centre in University Park, Pennsylvania. Los Alamos National Laboratory provides gamma-ray imaging analysis.



\bibliographystyle{mnras}
\bibliography{mylibrary}







\bsp
\label{lastpage}
\end{document}